\documentclass[11pt]{article}

\usepackage[T1]{fontenc}
\PassOptionsToPackage{pagebackref=true}{hyperref}
\usepackage{jheppub}
\usepackage{amsmath,braket,slashed,bm}
\usepackage{orcidlink}

\usepackage{cleveref}
\crefname{table}{Table}{Tables}
\crefname{equation}{Eq.}{Eqs.}
\crefname{appendix}{App.}{Apps.}
\crefname{section}{Sec.}{Secs.}
\crefname{figure}{Fig.}{Figs.}
\usepackage{enumitem}
\usepackage{tcolorbox}
\usepackage{cancel}
\newtcolorbox{condbox}{colback=gray!15,colframe=black!400,arc=2pt,left=1em,boxrule=0.5pt,top=3pt,bottom=3pt}
\usepackage{ytableau}
\usepackage{tikz}
\usepackage{tikz-feynman}
\tikzfeynmanset{compat=1.1.0}
\usetikzlibrary{positioning,calc,decorations.markings}
\usetikzlibrary{decorations.pathmorphing,arrows.meta}

\usepackage{dsfont}
\usepackage{subcaption}
\usepackage{graphicx}
\usepackage[normalem]{ulem}

\newcommand{\adj}{\text{adj}}
\newcommand{\RGB}{\mathbf{R}_{\mathrm{GB}}}

\def\vev#1{\left \langle #1 \right \rangle}
\DeclareMathOperator{\tr}{Tr}
\newcommand{\nn}{\nonumber}

\def\DALP{DALP}
\newcommand{\I}{\text{Inv}}

\def\MS{M_S}
\def\MF{M_F}
\def\lam{\mu}

\usepackage{colortbl}
\definecolor{mygreen}{rgb}{0.0, 0.5, 0.0}

\preprint{
\textcolor{blue}{IFT-UAM/CSIC-15-116\\
\,\mbox{} \hfill CERN-TH-2026-122}
}
\title{Darkly Charged ALPs}

\author[a]{Nicol\'as M. Arenaza\orcidlink{0009-0001-2693-1079},}
\author[a]{Enrique Fern\'andez-Mart\'inez\orcidlink{0000-0002-6274-4473},}
\author[a]{Bel\'en Gavela\orcidlink{0000-0002-2321-9190},}
\author[b]{Elizabeth E.~Jenkins\orcidlink{0000-0003-1404-5931},}
\author[b]{Aneesh V.~Manohar\orcidlink{0009-0004-5497-8554},}
\author[b,c]{Pablo Qu\'ilez\orcidlink{0000-0002-4327-2706},}
\author[a]{ and Thomas Steingasser\orcidlink{0000-0002-1726-2117}}

\affiliation[a]{
Departamento de F\'isica Te\'orica and Instituto de F\'{\i}sica Te\'orica, IFT-UAM/CSIC,\\
Universidad Aut\'onoma de Madrid, Cantoblanco, 28049, Madrid, Spain}
\affiliation[b]{Department of Physics, University of California, San Diego, La Jolla, CA 92093, USA}
\affiliation[c]{
CERN, Theoretical Physics Department, Geneva, Switzerland}

\abstract{
The established $d=5$ ALP effective Lagrangian describes the interaction of scalars with approximate shift-symmetry which carry no Standard Model (SM) charges with SM fields. It implicitly assumes that ALPs are not charged under any symmetries of the dark sector.  In this paper, we remove this assumption.  For ALPs carrying conserved dark charges,  no $d=5$ ALP effective interaction to SM particles is possible. We build the effective Lagrangian for these darkly charged ALPs stemming from a general breaking pattern, and we show that the lowest-order shift-symmetric effective Lagrangian contains just two $d=6$ operators coupling ALPs to SM particles. We explore the model-independent phenomenological implications of these interactions, as well as the question of whether the dark matter observed in the Universe may consist of darkly charged ALPs. We identify higher order operators of the effective field theory, and determine which types of dark symmetry groups can seed darkly charged ALPs. Illustrative examples of ultraviolet completions which result in darkly charged ALPs at low-energies are provided as well.  The darkly-charged ALP scenario can be generalized by including dark gauge interactions. In this paper, we have considered only the case with no such interactions.
}

\begin{document} 
\maketitle


\section{Introduction} \label{Sec:intro}
\noindent

Nature exhibits a variety of hidden (spontaneously broken) symmetries, whose tell-tale signals are massless or nearly massless (pseudo)-Goldstone bosons ((p)GBs), or the longitudinal components of massive gauge bosons. Examples include the longitudinal components of the electroweak $W^\pm$ and $Z$ gauge bosons, which are exact GBs of the $SU(3)_c\times SU(2)_L\times U(1)_Y \to SU(3)_c \times U(1)_Q$ gauge symmetry breakdown of the Standard Model (SM), and the light pions, whose lightness compared to $\Lambda_{\rm QCD}$ is understood in terms of their pGB nature from the breakdown of the approximate global chiral symmetry of QCD to its vector subgroup.   

It is pertinent to ask whether Nature  possesses other hidden symmetries which can account for outstanding shortcomings of the SM, such as the need to justify surprisingly small parameters or the lack of dark matter (DM) candidates which explain the cosmological and astrophysical data. This quest includes scenarios which solve the strong CP problem~\cite{Peccei:1977hh,Peccei:1977ur,Weinberg:1977ma,Wilczek:1977pj}, theories with extra space-time dimensions~\cite{ArkaniHamed:1998pf,Dienes:1999gw,Chang:1999si,DiLella:2000dn}, some dynamical flavor theories~\cite{Davidson:1981zd,Wilczek:1982rv,Ema:2016ops,Calibbi:2016hwq}, dark matter models~\cite{Abbott:1982af, Dine:1982ah,Preskill:1982cy}, scenarios which explore a dynamical origin for Majorana neutrino masses~\cite{Gelmini:1980re} and string-inspired  models~\cite{Cicoli:2013ana}, among others. The most prominent pGB candidate is the QCD axion~\cite{Peccei:1977hh,Peccei:1977ur,Weinberg:1977ma,Wilczek:1977pj,Kim:1984pt,Choi:1985cb,Zhitnitsky:1980tq,Dine:1981rt,Kim:1979if,Shifman:1979if}, which aims to explain the absence of CP violation in the strong sector by dynamically relaxing the $\bar \theta$-parameter to the CP conserving point; the small axion mass stems then from its anomalous couplings to QCD.  

Nowadays, the term axion-like particle (ALP) refers to any pGB {\it a priori} unrelated to the strong CP problem, with a generic small mass of undefined origin.  The dynamics of ALPs is formulated in terms of effective couplings, which do not depend on any particular ultraviolet-complete model.  The present extremely intense search for ALPs constitutes a phenomenal quest  to uncover fundamental symmetries hidden in Nature which can be awaiting discovery.   More precisely, ALPs are generic pGBs which are singlets of the SM gauge group, stemming from the spontaneous breakdown of a (yet undiscovered) global symmetry.  The lowest order (LO) effective couplings of the customary ALP Lagrangian have mass-dimension five  $(d=5)$~\cite{Georgi:1986df} and are linear in  $a/f_a$, where $a$ is  the ALP field, and  $f_a$, the ALP symmetry-breaking scale, is assumed to be larger than the electroweak scale $v$,  $f_a \gg v$.  The setup is compatible with the assumption that the ALP field is the pGB  of a  spontaneously  broken $U(1)$ symmetry.
 A vigorous and world-wide surge in theoretical and experimental efforts to tackle this type of ALP is unfolding.  However, this ALP Lagrangian carries an implicit and unfounded assumption, namely that the ALP is a singlet not only of the SM gauge group but also of any possible symmetry of the dark sector of the Universe.  In other words, it assumes that the ALP carries no dark charge whatsoever.   This is  also true, in general, in scenarios with multiple axions: these setups are typically consistent with assuming that the spontaneously broken dark symmetry corresponds to  pure $U(1)$ factors~\cite{Gavela:2023tzu,Chadha-Day:2021uyt,deGiorgi:2025ldc,Murai:2023xjn,Murai:2024nsp,Dunsky:2025sgz,Chadha-Day:2023wub,Baryakhtar:2026oun,FernandezNavarro:2026cyu,Dessert:2025yvk,Grinstein:2026alk}, or in any case that the dark symmetry  has been completely broken.  

In this paper, we remove this assumption and explore the effective model-independent couplings of ALPs which may be charged under some exactly conserved dark symmetries. The SM fields by definition are assumed to be neutral with respect to the dark charges. Dark charge conservation then requires at least two ALP fields participating in any effective operator,  irrespective of the type of dark charge.  This requirement automatically forbids the entire $d=5$ ALP Lagrangian which is linear in the ALP field; the leading order (LO) effective field theory (EFT) will be shown to start at dimension six, and have two possible $d=6$ couplings.  We will denote in what follows the generic darkly-charged pGBs or ALPs as DALPs, and we focus on their shift-invariant couplings, which respect the non-linear symmetry of the spontaneous symmetry breaking pattern $G \to H$ in the dark sector.
 The ensuing phenomenology differs substantially from that for customary uncharged ALPs, whose leading interactions are forbidden for DALPs, including the widely explored ALP-photon-photon coupling.  Conversely, other couplings come to the forefront.
 
 We explore a new ALP coupling of the form $B_{\mu \nu}( \partial_\mu a_1)( \partial_\nu a_2)$ which generically exists in theories with more than one ALP. It does not exist, though, in the most commonly explored ALP scenarios in the literature: $U(1)$ breaking or  dark pions from QCD-like breakings~\cite{Bhattacharya:2013kma,Hochberg:2014dra,Alexander:2023wgk,Garcia-Cely:2025flv,Chu:2024rrv}. However, generic  symmetry breaking scenarios do generate such a coupling.\footnote{For scenarios with several spontaneously broken $U(1)$'s, this operator is present~\cite{Dessert:2025yvk} but it is subleading, as the phenomenology is typically dominated by the allowed $d=5$ couplings.}
 
We consider the general case where the ALPs take values in a coset $G/H$, and give the conditions for the existence of the $B_{\mu \nu}$ interaction.  These theories are of three main types: (a) flat  cosets from the breaking of $U(1)$, (b) symmetric cosets, and (c) non-symmetric cosets. This nomenclature, and the differences between the three cases, will be explained in \cref{sec:chpt}. The symmetry $G$ may have small symmetry breaking terms, so the DALPs are light compared with the symmetry breaking scale, but not exactly massless. It is possible that a subgroup of $G$ is gauged. In this case, the GBs corresponding to the broken gauged generators lead to massive dark gauge bosons, rather than GBs. Models of this type have been considered for technicolor theories, where the vacuum alignment of $H$ in $G$ has been studied~\cite{Preskill:1980mz,Peskin:1980gc}. In this paper, we focus on examples where the symmetry group $G$ is purely global.  Generalization to the case of gauged dark symmetries is possible, and it becomes straightforward when only a subgroup of $H$ is gauged.

 The DALP scenario and its leading-order interaction Lagrangian are summarized in \cref{sec:DALPscenario}. Readers not interested in the technical details of $G/H$ sigma models can skip directly to \cref{sec:pheno}, which depends only on the summary overview given in \cref{sec:DALPscenario}.  The general structure of DALP interactions and the derivation of the DALP EFT Lagrangian are given in \cref{sec:chpt}.  Two group theoretic conditions which result in the DALP interaction Lagrangian are identified, and numerous examples of $G \to H$ symmetry breakings which satisfy these conditions are presented.  Section~\ref{sec:pheno} addresses experimental constraints and signals of the leading DALP EFT.   Several ultraviolet (UV) complete models are considered in \cref{sec:uvcompletions}. Conclusions are presented in \cref{sec:conclusions}.
   
\section{The Dark Chiral Lagrangian: Summary}\label{sec:DALPscenario}

We consider a scenario with multiple ALPs which are (pseudo)-Goldstone bosons of a spontaneously broken global symmetry $G\to H$ at a scale $f_a$, with the SM fields uncharged under $G$ and the ALPs uncharged under the SM gauge group. These ALPs are assumed to carry a conserved dark charge,\footnote{ A theory with a single ALP field  can carry at most a $\mathbb{Z}_2$ charge as it is a real scalar; this case has been studied in Refs.~\cite{Weinberg:2013kea,Bauer:2022rwf}. Larger discrete symmetries have been studied in Ref.~\cite{Vileta:2022jou}; we focus on continuous $H$.  } and thus they must transform as a non-trivial representation $\RGB$ of the unbroken group $H$. All the customary $d=5$ ALP operators, such as $\partial_\mu a\, j^\mu_{\text{SM}}$ and $a\, X^{\mu \nu} \widetilde{X}_{\mu \nu}$, where $X_{\mu \nu}$ is a gauge field-strength tensor, are then forbidden, as they are linear in the ALP fields and therefore not $H$-invariant.

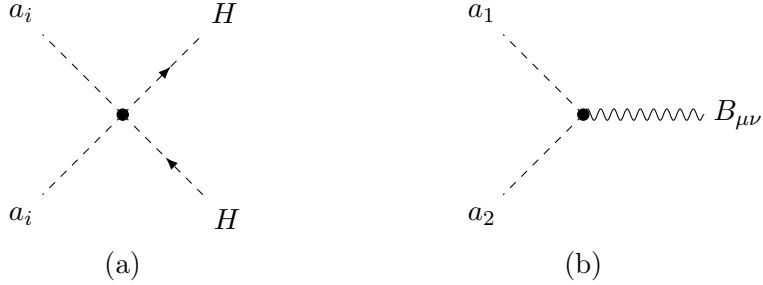
\begin{figure}[h]
\begin{center}
\tikzset{photon/.style={decorate, decoration={snake, segment length=1.75mm, amplitude=0.75mm}}}
%
\begin{tikzpicture}
\coordinate (v) at (0,0);
\draw[dashed] (v) -- +(135:1.5);
\draw[dashed] (v) -- +(225:1.5);
\draw[dashed,-Latex] (v) -- +(45:0.9);
\draw[dashed] (v)+(45:0.9) -- +(45:1.5);
\draw[dashed] (v) -- +(-45:0.8);
\draw[dashed,-Latex] (v)+(-45:1.5) -- +(-45:0.8);
\filldraw (v) circle (0.075);
\draw (v)+(135:1.9) node {$a_i$};
\draw (v)+(225:1.9) node {$a_i$};
\draw (v)+(45:1.9) node {$H$};
\draw (v)+(-45:1.95) node {$H$};
\draw (0,-2) node {(a)};
\end{tikzpicture}
\hspace{2.5cm}
\begin{tikzpicture}
\coordinate (v) at (0,0);
\draw[dashed] (v) -- +(135:1.5);
\draw[dashed] (v) -- +(225:1.5);
\draw[photon] (v) -- +(0:1.6);
\filldraw (v) circle (0.075);
\draw (v)+(135:1.9) node {$a_1$};
\draw (v)+(225:1.9) node {$a_2$};
\draw (v)+(0:2.05) node {$B_{\mu\nu}$};
\draw (0,-2) node {(b)};
\end{tikzpicture}
\end{center}
\caption{\label{fig:LOoperators} Feynman diagrams of the two LO DALP operators: (a) the universal Higgs-portal operator $\mathcal{O}_{Ha}$ and (b) the antisymmetric operator $\mathcal{O}_{Ba}$.}
\end{figure}

The ALP fields will be denoted by $a_i$, $i=1,\cdots, n_\text{ALP}$ and are real scalar fields; a complex scalar field can always be written as a linear combination of two real scalar fields. The case we consider, with darkly-charged ALPs, has $n_\text{ALP} > 1$. The leading interactions of darkly charged ALPs with the SM fields are $d=6$. There are only two possible types of operators, 
\begin{equation}
\mathcal{L}_{\rm{LO}}^{d=6}= \frac{1}{f_a^2}\, S_{ij}\, (H^\dagger H) \left\{ (\partial_\mu a_i)( \partial^\mu a_j)   \right\} \,+\,  \frac{1}{f_a^2}\, A_{ij} 
B^{\mu \nu} \left\{ (\partial_\mu a_i)( \partial_\nu a_j)  \right\}\,,
\label{eq:DALP-LO-Lag1}
\end{equation}
where $S_{ij}$ is a symmetric $H$-invariant tensor,  $A_{ij}$ is an antisymmetric $H$-invariant tensor, $B_{\mu \nu}$ is the hypercharge field-strength\footnote{One can wonder whether there is another $d=6$ operator similar to the second one in Eq.~(\ref{eq:DALP-LO-Lag1}), albeit with $B_{\mu \nu}$ replaced by $\widetilde B_{\mu \nu}$: it vanishes identically though, on integration by parts.  Other couplings containing $\widetilde B_{\mu \nu}$  are possible  and  discussed further around \cref{3.23}.} and a sum over repeated indices is understood in this paper. The interactions they seed are illustrated in Fig.~\ref{fig:LOoperators}.
The first operator is a Higgs portal which is universal and exists for any dark symmetry group, since there is a symmetric invariant tensor $\delta_{ij}$ that enters the ALP kinetic energy term
\begin{align}
\mathcal{L}_\text{kin} &= \frac12 \delta_{ij}\left\{ ( \partial_\mu a_i )( \partial^\mu a_j) \right\} \,,
\end{align}
and thus $S_{ij} = c_{Ha} \delta_{ij}$ is a possible interaction in \cref{eq:DALP-LO-Lag1}, where $c_{Ha}$ is the Higgs-portal coupling~\cite{Weinberg:2013kea,Bauer:2022rwf}. 
The second operator in \cref{eq:DALP-LO-Lag1} is a hypercharge portal: it requires an antisymmetric invariant, since $B_{\mu \nu}$ is antisymmetric in its indices. In this paper, we will later consider models where such an invariant exists. General symmetry breaking patterns $G \to H$ provide additional terms to  \cref{eq:DALP-LO-Lag1} which are required by the non-linear symmetry; these terms contain more powers of $a_i/f_a$, are fixed by symmetry, and have mass dimension $d=7,8,\cdots$.  They are given explicitly in \cref{sec:chpt}.

To simplify the analysis of experimental constraints, we focus in \cref{sec:pheno} on a minimal DALP scenario with $n_\text{ALP}=2$. In this case, there is only one symmetric invariant $\delta_{ij}$, and the antisymmetric invariant is proportional to $\epsilon_{ij}$, so that \cref{eq:DALP-LO-Lag1} reduces to
\begin{equation}
\mathcal{L}_{\rm{LO}}^{d=6}= \frac{c_{Ha}}{f_a^2}\,   \mathcal{O}_{Ha}  \,+\,  \frac{c_{Ba}}{f_a^2}\,  \mathcal{O}_{Ba},
\label{eq:DALP-LO-Lag}
\end{equation}
where
\begin{align}
\mathcal{O}_{Ha} &=  (H^\dagger H) (\partial_\mu a_i)( \partial^\mu a_i) = (H^\dagger H)\left[  (\partial_\mu a_1)( \partial^\mu a_1)+ (\partial_\mu a_2)( \partial^\mu a_2)\right] \,, \nn \\
\mathcal{O}_{Ba} &=  \epsilon_{ij} B^{\mu \nu} (\partial_\mu a_i)( \partial_\nu a_j) = 2 B^{\mu \nu} (\partial_\mu a_1)( \partial_\nu a_2)\,,
\label{2.4}
\end{align}
and $c_{Ha}$ and $c_{Ba}$ are coupling constants.  
$\mathcal{O}_{Ba}$ is the only possible $d=6$ DALP operator  which includes a SM field strength, i.e. $B_{\mu \nu}$, since the $W$ and gluon field-strengths transform nontrivially under the SM gauge group.
 
The $G$ symmetry can be weakly broken, in which case the darkly charged ALPs are not exactly massless Goldstone bosons, but rather pseudo-Goldstone bosons with masses much lighter than the axion decay constant, $m_i\ll f_a$.   In consequence, the \DALP\ EFT is
\begin{equation}
\mathcal{L_{\rm{eff}}} = \mathcal{L_{\rm{SM}}} \,+\, 
\frac12\left(\partial_\mu a_i\, \partial^\mu a_i - m_i^2 a_i^2\right)\,+\, \mathcal{L}_{\rm{LO}}^{d=6}  +\dots
\label{2.20}
 \end{equation}
at LO ($d=6$) in the EFT expansion.  The structure of this spectrum, and in particular the degeneracy of DALPs within each $H$-multiplet whenever the $G$-breaking terms respect $H$, is discussed in \cref{subsec:dalp_masses}.

The next section provides a detailed derivation of the DALP EFT and the $d=6$ interactions, and also the structure of higher-dimension terms. Readers not interested in the technical details can assume  \cref{eq:DALP-LO-Lag} and skip directly to the phenomenological constraints in \cref{sec:pheno}.

\section{The Dark Chiral Lagrangian}\label{sec:chpt}

In this section, we construct the DALP EFT for a general $G \to H$ symmetry breaking pattern using the CCWZ formalism~\cite{Coleman:1969sm,Callan:1969sn}, and give concrete examples which lead to the DALP EFT Lagrangian \cref{eq:DALP-LO-Lag}. We will use the familiar terminology from chiral perturbation theory, even though the dark symmetry $G$ has nothing to do with QCD chiral symmetry. The DALP Lagrangian is constructed from the chiral field $u_i^\mu = \partial^\mu a_i/f_a + \cdots$, where the dots denote terms with additional powers of $a_i/f_a$, which are of the same order in momentum (i.e.\ order $p/f_a$) but of higher mass dimension in the EFT power counting. These higher-dimension  terms inside  $u_i^\mu$ are fixed by the non-linear realization of $G$-symmetry, and are derived later in this section. The field $u_i^\mu$ is covariant under the full symmetry group $G$. The usual shift symmetry is the transformation rule under broken symmetry transformations to lowest order in $a_i/f_a$.

Under the symmetry breaking $G \to H$, the generators of $G$, which are in the adjoint representation of $G$, transform under $H$ according to the branching rule
\begin{align}
\adj\, G \Big|_H & = \adj\, H \oplus \RGB \,.
\label{1.1}
\end{align}
The generators transforming as $\adj\, H$ are the unbroken generators. The remaining generators, which are broken, transform as $\RGB$ under $H$, and are the representation of the GBs. The representation $\RGB$ can be reducible. For example, under the breaking $SU(3) \to SU(2) \times U(1) $ with the regular embedding, $\mathbf{8} \to \mathbf{3}_0  \oplus \mathbf{1}_0 \oplus  \mathbf{2}_1 \oplus \mathbf{2}_{-1} $, and $\RGB = \mathbf{2}_{1} \oplus \mathbf{2}_{-1}$. The number of ALPs, which is the number of broken generators, is $n_\text{ALP} = \dim \RGB = \dim G - \dim H $.

For a single ALP arising from the breaking $U(1)\to \varnothing$~\cite{Georgi:1986df,Kaplan:1985dv}, or more generally when $G$ is completely broken, the unbroken group $H$ is trivial and  $d=5$ operators linear in the ALP field are present. The more general symmetry breaking pattern of the dark sector $G\to H$ with non-trivial $H$ forbids the usual $d=5$ operators of a single ALP field coupled to SM particles by $H$-invariance. The condition for this is
\begin{itemize}[leftmargin=4em]
\item[\bf{[C1]:}] $\RGB\not\supset \I_H$: there are no $H$-invariants in $\RGB$, which ensures that $d=5$ operators linear in ALP fields are absent.
\end{itemize}

We consider models which have the $B_{\mu \nu}$ operator in \cref{eq:DALP-LO-Lag}. The condition for this is
\begin{itemize}[leftmargin=4em]
\item[\bf{[C2]:}] $(\RGB \otimes \RGB)_A \supset \I_H$: there is an $H$-invariant that is an antisymmetric bilinear in the GB fields, so that a non-vanishing $\mathcal{O}_{Ba}$ is allowed. Note that this condition requires a minimum of two ALPs.
\end{itemize}
These two conditions are reformulated purely in terms of general properties of the groups $G$ and $H$, without any reference to the Goldstone representation $\RGB$, later in this section. Many $G \to H$ symmetry breaking examples which satisfy the two conditions are presented.

\Cref{sec:ccwz} gives a brief review of the construction of EFT for the interactions of GBs.  \Cref{sec:darkchirallagrangian} constructs the chiral Lagrangian of dark ALPs which carry non-trivial $H$ quantum numbers at LO in the chiral expansion.  \Cref{subsec:conditions} explores the consequences of the two conditions on the GB representation $\RGB$.  Many concrete examples illustrating the conditions are presented in \cref{sec:examples}.

\subsection{The CCWZ construction}\label{sec:ccwz}

The general formalism for spontaneously broken theories was worked out by Callan, Coleman, Wess and Zumino (CCWZ)~\cite{Coleman:1969sm,Callan:1969sn}. We briefly summarize their results, which we need to classify the EFT operators required in our analysis.

The generators of the symmetry group $G$ are $t_A$, which divide into unbroken generators $T_\alpha$ of $H$, and broken generators $X_i$, $\{t_A\} = \{T_\alpha, X_i\}$. The $G$ generators $t_A$ transform as the adjoint representation of $G$; under the unbroken subgroup $H$, this adjoint representation splits as in \cref{1.1}. The decomposition into $T_\alpha$ and $X_i$ may not be unique. If $\RGB$ contains $\text{adj}\, H$, as it happens in QCD, one can shift some of the broken generators by the unbroken generators $T_\alpha$. Different choices of the broken generators simply correspond to choices of basis and lead to effective Lagrangians related by field redefinitions. We fix some choice of broken generators for our analysis.

For each representation $R$, the normalization of the generators of $G$ can be expressed through its Dynkin index $T(R)$ as
\begin{equation}
\mathrm{Tr}(T_\alpha T_\beta)=T(R)\, \delta_{\alpha\beta}, \qquad
\mathrm{Tr}(X_i X_j)=T(R)\, \delta_{ij}, \qquad
\mathrm{Tr}(T_\alpha X_i)=0.
\label{3.2a}
\end{equation}
This allows to express the coefficients in the expansion of any element $\phi$ of the Lie algebra  in terms of the generators as
\begin{align}
    \bm{\phi} &= \alpha_i X_i + \beta_a T_a & & \Leftrightarrow & \alpha_i & = \frac{\mathrm{Tr} (\bm{\phi} X_i)}{T(R)} , &   \beta_a & = \frac{\mathrm{Tr} (\bm{\phi} T_a)}{T(R)}\,.
\label{3.3a}
\end{align}
Usually, the generators of $G$ which define the GB fields are chosen to be in the fundamental representation, with $T(R)=1/2$.

The commutator relations of the generators of $G$ are
\begin{align}
[t_A,t_B] &= i f_{ABC} t_C\,,
\label{1.16a}
\end{align}
where the structure constants $f_{ABC}$ are real and completely antisymmetric. The structure constants $f_{ABC}$ with indices restricted to broken and/or unbroken generators will be denoted by $f_{\alpha\beta\gamma}$, $f_{\alpha\beta i}$, $f_{\alpha ij}$ and $f_{ijk}$, i.e.\ lower case greek (latin) indices for unbroken (broken) generators.
The commutator of two unbroken generators is an unbroken generator,
\begin{align}
[T_\alpha,T_\beta] &= i f_{\alpha\beta\gamma} T_\gamma\,,
\label{1.16}
\end{align}
and the commutator of an unbroken generator with a broken generator is a broken generator, since the broken generators transform as $\RGB$ under $H$,
\begin{align}
[T_\alpha,X_i] &= i f_{\alpha ij} X_j\,.
\label{1.17}
\end{align}
In general, the commutator of two broken generators
\begin{align}
[X_i,X_j] &= i f_{ij\alpha} T_\alpha + i f_{ijk} X_k
\label{1.18}
\end{align}
contains both unbroken and broken generators. A \emph{symmetric coset} is one where the broken generators can be chosen so that the Lie algebra has a symmetry $T_\alpha \to T_\alpha$, $X_i \to -X_i$, so that the commutator of two broken generators is an unbroken generator and $f_{ijk}$ vanishes. We give examples with both symmetric and non-symmetric cosets in \cref{sec:examples}.

The field $\xi(x)$ in the CCWZ formalism is defined as
\begin{align}
\xi(x) &= e^{ i\bm{\theta}(x)} \,, & \bm{\theta} &= \theta_i(x) X_i= \frac{a_i(x)}{f_{a,i}}X_i \,,
\label{1.2}
\end{align}
which is the exponential of the broken generators, and describes the local orientation of the vacuum of the theory. It parameterizes the cosets $G/H$. The fields $\theta_i(x)$ are dimensionless Goldstone boson fields, and are the analog of $\pi_i(x)/f_\pi$ in QCD. After writing down the EFT kinetic energy term, we will introduce the symmetry breaking scale $f_a$ and switch to conventional scalar fields $a_i$ with mass dimension one. 

The transformation law for $\xi(x)$ under a \emph{global} $G$ transformation is
\begin{align}
\xi(x) &\to \xi^\prime(x) = g\, \xi(x) h(x)^{-1}
\label{1.3}
\end{align}
where $g \in G$ and $h \in H$.  The transformation $h(x)$ depends on $x$ through the $x$-dependence of the Goldstone boson fields $\theta_i(x)$. Under a \emph{global} $H$ transformation $h_0$,
\begin{align}
\bm{\theta}^\prime(x) &=h_0\, \bm{\theta}(x) h_0^{-1}\,, & 
h(x) &= h_0 \,.
\label{1.4}
\end{align}
Under the unbroken group, $\bm{\theta}$ transforms linearly in the same way as the broken generators, i.e.\ as representation $\RGB$.

Goldstone bosons are derivatively coupled, and the basic object used to construct the chiral Lagrangian is the Maurer-Cartan form,
\begin{align}
\xi^{-1} (x) \partial^\mu \xi (x) &=  i u^\mu(x) + i \Gamma^\mu(x)\,, &
u^\mu(x) &= u_i^\mu(x) X_i \,, &
\Gamma^\mu(x) &= \Gamma_\alpha^\mu(x) T_\alpha \,.
\label{1.7}
\end{align}
The l.h.s.\ $\xi^{-1} \partial^\mu \xi$ can be expanded in terms of multiple commutators of $\bm{\theta}$ with $\partial_\mu \bm{\theta}$, 
\begin{equation}
  \xi^{-1}\partial_\mu\xi \;=\; i\,\partial^\mu\boldsymbol{\theta} \;+\; \tfrac{1}{2}\,[\boldsymbol{\theta},\,\partial^\mu\boldsymbol{\theta}] \;-\; \tfrac{i}{6}\,[\boldsymbol{\theta},\,[\boldsymbol{\theta},\,\partial^\mu\boldsymbol{\theta}]] \;+\; \cdots\,,
  \label{1.7a}
\end{equation}
and it is thus a linear combination of the generators of $G$, which splits into a linear combination of broken generators $u^\mu$ and a linear combination of unbroken generators $\Gamma^\mu$. Expanding in powers of $\bm{\theta}$, one obtains
\begin{align}
u_i^\mu(x) &= \partial^\mu \theta_i(x) + \ldots \equiv e_{ij}(\theta) \partial^\mu \theta_j(x), \nn \\
\Gamma^\mu_\alpha (x) &= \frac12 f_{ij\alpha} \theta_i \partial^\mu \theta_j + \ldots\,,
\label{1.9}
\end{align}
using \cref{3.2a} and \cref{3.3a} to extract $u^\mu_{\,i}$ and $\Gamma^{\mu}_\alpha$ by taking traces.  The dots in \cref{1.9} denote other terms of first order (i.e.\ order $p$) in the chiral expansion but higher order in the EFT mass-dimension power counting. These higher order terms  are required by the $G$ symmetry, which is implemented nonlinearly in the chiral theory.  The coefficients $e_{ij}(\theta) = \delta_{ij} + \ldots$ are vielbeins on the coset space $G/H$. If $G/H$ is a flat coset, $e_{ij} (\theta) = \delta_{ij}$, then $u^\mu_i = \partial^\mu \theta_i$, with no higher order terms. This special case occurs for the breaking of a $U(1)$ factor, since $U(1)$ is a circle, which has zero curvature. For a symmetric coset, $e_{ij}(\theta)$ contains only even powers of $\theta$, so $u_i^\mu$ only contains odd powers of $\theta$ and  the dots start at  $\mathcal{O}(\theta^3)$. For a non-symmetric coset, $e_{ij}(\theta)$  and $u_i^\mu$ have all powers of $\theta$,
\begin{equation}
u^\mu_{\,i}
= \partial^\mu \theta_i
+ \frac{1}{2} f_{jki}\,\theta_j \partial^\mu \theta_k
+ \frac{1}{6} f_{jkA}{} f_{lAi}\,
\theta_l \theta_j \partial^\mu \theta_k
+ \mathcal{O}(\theta^4).
\label{1.9a}
\end{equation}
For a symmetric coset, the corresponding expression is \cref{1.9a} with $f_{jki} = 0$. For any coset, the expansion of the connection $\Gamma^\mu_{\,\alpha}$ up to cubic terms reads
\begin{equation}
\Gamma^\mu_{\,\alpha}
= \frac{1}{2} f_{ij\alpha}\,\theta_i \partial^\mu \theta_j
+ \frac{1}{6} f_{ijA} f_{kA\alpha}{}\,
\theta_k \theta_i \partial^\mu \theta_j
+ \mathcal{O}(\theta^4).
\end{equation}

Under the chiral transformation \cref{1.3},
\begin{align}
u^\mu(x) &\to h(x)\, u^\mu(x) \, h(x)^{-1} \, ,\nn \\
\Gamma^\mu(x) &\to h(x)\, \Gamma^\mu(x) \, h(x)^{-1} + i \partial^\mu h(x) \, h(x)^{-1} \, .
\label{1.11}
\end{align}
$u_i^\mu(x)$ transforms as $\RGB$, and $\Gamma_\alpha^\mu(x)$ transforms as an $H$ gauge field.
The chiral covariant derivative $\nabla_\mu$ is defined as
\begin{align}
\nabla_\mu &= \partial_\mu + i \Gamma_\mu, & 
\nabla^\mu u^\nu &\equiv \partial^\mu u^\nu + i \left[ \Gamma^\mu, u^\nu \right] \, .
\label{1.12}
\end{align}

We now have the building blocks to construct the EFT Lagrangian for the interactions of darkly charged ALPs with Standard Model fields. The field $u^\mu$ is order $p$ in the chiral counting, and starts at operator dimension two in the EFT power counting~\cite{Manohar:1983md,Buchalla:2013eza,Gavela:2016bzc, Brivio:2025yrr}, since $\partial^\mu \theta_i = \partial^\mu a_i/f_a$. The factor $1/f_a \sim 4\pi/\Lambda$ is the inverse factor of $\Lambda$ required by the EFT power counting. Each additional power of $\theta$ in the expansion of $u^\mu$ increases the field dimension by one. The EFT Lagrangian is constructed by combining chiral operators which are $H$-invariant (and hence $G$-invariant under the non-linear transformation law \cref{1.3}) with Standard Model operators.

\subsection{Leading order dark chiral Lagrangian}\label{sec:darkchirallagrangian}

The first condition {\bf{[C1]}} implies that all DALPs carry non-trivial $H$ quantum numbers, and so linear terms in $\theta_i$ are not $H$-invariant and are forbidden in the EFT Lagrangian.  Consequently, the lowest order chiral invariant operator in the EFT is order $p^2$ in the chiral counting and mass dimension four,
\begin{align}
S_{ij} u^\mu_i\, u_{j\mu} &= S_{ij}\, \partial^\mu \theta_i \, \partial_\mu \theta_j + \ldots \, ,
\label{1.14}
\end{align}
where $S_{ij}$ is a symmetric tensor which is $H$-invariant and independent of $\theta$.  One symmetric invariant always exists, $S_{ij}\propto\delta_{ij}$, but there can be others. This result is explained further with the help of explicit examples in \cref{sec:examples}. Let us choose $S_{ij}$ to have mass dimension two, e.g. $S_{ij} = f_{a,i}^2 \delta_{ij}$. The Goldstone boson kinetic term arises from
\begin{align}
\mathcal{L} &= \frac12  S_{ij} u^\mu_i\, u_{j\mu}  = \frac12 S_{ij} \, \partial^\mu \theta_i \, \partial_\mu \theta_j + \ldots = \frac12 \frac{S_{ij}}{f_{a,i} f_{a,j}} \partial^\mu a_i \partial_\mu a_j + \ldots = \frac12 \partial^\mu a_i \partial_\mu a_i + \ldots \, ,
\label{1.15}
\end{align}
where the Goldstone boson fields are $\theta_i = a_i/f_{a,i}$, and the DALP fields $a_i$, which are scalar fields, have mass dimension one. The decay constants $f_{a,i}$ are chosen so that the ALP kinetic energy is canonically normalized.\footnote{More precisely, first make an orthogonal transformation to diagonalize $S_{ij}$, and then determine $f_{a,i}$.} Not all Goldstone bosons need to have the same decay constant; e.g., in QCD, we have $f_K \not=f_\pi$. For symmetric cosets, the kinetic term \cref{1.15} has an expansion in even powers of $a_i$, and in general contains terms of mass dimension 4, 6, 8, etc. For non-symmetric cosets, it also contains terms which are odd powers of $a_i$, and contains terms of mass dimension 4, 5, 6, etc.

Higher order terms in the chiral EFT are built from more powers of $u_\mu$ and chiral covariant derivatives $\nabla_{\!\mu}$, and have been studied extensively for QCD~\cite{Gasser:1983yg,Bijnens:1999sh}. The first correction after the kinetic term is fourth order in $u_\mu$, and so it is order $p^4$ and starts at mass dimension eight. To mass dimension six, the only purely Goldstone boson operators are from the expansion \cref{1.15}.

We now consider the coupling of the dark sector to the Standard Model. Condition {\bf{[C1]}} implies that for DALPs carrying non-trivial dark charges,  the interaction must start at least at second order in $u_\mu$, amounting to  an operator of at least  mass dimension four. This operator is coupled with the lowest order SM singlet ``portals'' which turn out to have mass dimension two: i.e. $H^\dagger H$ where $H$ denotes the SM Higgs doublet, and the hypercharge field strength $B_{\mu\nu}$.  This already suggests that  the DALP effective Lagrangian describing interactions between SM and DALPs will include terms of mass dimension six $(d=6)$, as we expatiate next.

There is a universal term of order $p^2$ and starting at $d=6$:
\begin{align}
\widetilde S_{ij} u_\mu^i\, u^{j\mu} \, (H^\dagger H) \, ,
\label{1.19}
\end{align}
where $\widetilde S_{ij}$ is a symmetric invariant tensor. Note that $\widetilde S_{ij}$ can differ from $S_{ij}$ in the kinetic energy term \cref{1.15}. The second condition {\bf{[C2]}} requires the existence of a $H$-invariant two-index antisymmetric tensor $A_{ij}$. There may be more than one antisymmetric tensor $A_{ij}$; we assume there is at least one. Then, for a given antisymmetric tensor $A_{ij}$, there are two more operators
\begin{align}
 A_{ij} u^\mu_i\, u^\nu_j \, B_{\mu \nu}, \qquad  A_{ij} u^\mu_i\, u^\nu_j \, \widetilde B_{\mu \nu} \, ,
\label{1.20}
\end{align}
which both are order $p^2$ and naively start at mass dimension six. The second one actually starts at $d=7$ in the EFT expansion; see discussion around \cref{3.23}.

In summary, the lowest dimension couplings of the dark ALP sector to the Standard Model start at mass dimension six. The higher order in $a_i$ terms from \cref{1.19} and \cref{1.20} are dimension eight for symmetric cosets and dimension seven for non-symmetric cosets.

One can systematically classify the interactions of dark ALPs with Standard Model fields. The general way to implement the chiral symmetry is to write down the most general interaction Lagrangian of the $G/H$ sigma model with the SM, using $u^\mu$ and $\nabla_\mu$ for the sigma model sector, and make field redefinitions which preserve the chiral and SM transformation properties of the fields. Then the SM fields remain invariant under chiral transformations, and the chiral field remains invariant under SM transformations.\footnote{An example of a transformation not of this form is $\psi \to e^{ia \gamma_5/f_a} \psi$ which turns the ALP coupling from axial vector to pseudoscalar. The new fermion field transforms under the shift symmetry. One is free to make additional field redefinitions of this type after determining the allowed operators in \cref{tab:operators}.}

The chiral operators are  constructed from  powers of $u_\mu$ and its covariant derivatives. The chiral equation of motion eliminates $\nabla_\mu u^\mu$. The identity $\nabla_\mu u^i_\nu - \nabla_\nu u^i_\mu = f_{ijk} u^j_\mu u^k_\nu$ eliminates the antisymmetric combination $\nabla_\mu u^i_\nu - \nabla_\nu u^i_\mu$. The resulting operators have the form $O^{\mu_1 \cdots \mu_n} _\chi O^{\mu_1 \cdots \mu_n} _\text{SM}$, where $O_\chi$ is an $H$-invariant chiral operator, and $O_\text{SM}$ is a  gauge invariant SM operator. Total derivatives on $O_\chi$ can be transferred to $O_\text{SM}$ by integration by parts. A straightforward analysis shows that the lowest dimension interactions are those shown in Table~\ref{tab:operators}, which classifies the couplings of pGBs to SM fields including both the case of pGBs charged  under some dark symmetry  as in the scenario considered in this paper (i.e. DALPs), or uncharged (as in the usual $U(1)$ ALP Lagrangian).   Operators in the $\dim \ge 5$  row of Table~\ref{tab:operators}  are shown for completeness even though they are forbidden in the DALP scenarios considered in this paper, since $\RGB$ does not contain any $H$-invariant; this includes higher-dimension operators linear in $u_\mu$ such as $ u_\mu^i (H^\dagger H)( \overline \psi \gamma^\mu \psi)$, etc. The operators in all other rows do exist {\it a priori} for DALPs. Some pertinent remarks on the latter follow:
 \begin{table}[t!]
\renewcommand{\arraystretch}{1.5}
\centering
    \begin{tabular}{|c|c|c|}
    \hline
    Dim & Operator Structure & Lowest Dimension Examples \\ \hline
    
    $\dim \ge 5$ & $u^i_\mu \, j^\mu_\text{SM}$ & $i u^i_\mu \left( H^\dagger D_\mu H - (D_\mu H)^\dagger H \right)$, $u^i_\mu \overline \psi \gamma^\mu \psi$ \\ \hline
    
    $\dim \ge 6$ &  $S_{ij}  u^i_\mu u^{j\mu} O_\text{SM}$, $S_{ij}  u^i_\mu u^j_\nu O_\text{SM}^{(\mu \nu)}$, $A_{ij}  u^i_\mu u^j_\nu O_\text{SM}^{[\mu \nu]}$ & $S_{ij}  u^i_\mu u^j_\mu (H^\dagger H), \ A_{ij}  u^i_\mu u^j_\nu B^{\mu \nu}$ \\ \hline

    $\dim \ge 7$ &  $A_{ij}  u^i_\mu u^j_\nu O_\text{SM}^{[\mu \nu]}$ with $\partial_\mu O_\text{SM}^{[\mu \nu]}=0$   &  $A_{ij}  u^i_\mu u^j_\nu \widetilde B^{\mu \nu}$ \\ \hline

    $\dim \ge 8$ & $A_{ij} \left[ (\nabla_\nu u^{ i\alpha}) u^j_\alpha -(\nabla_\nu u^{j\alpha}) u^i_\alpha \right]  j^\nu_\text{SM}$  &  $A_{ij} \left[ (\nabla_\nu u^{ i\alpha}) u^j_\alpha -(\nabla_\nu u^{j\alpha}) u^i_\alpha \right] \overline \psi \gamma^\nu \psi$ \\[3pt] \hline

     & $H_{ijk} u^i_\mu u^{j\mu} u^k_\nu j_\text{SM}^\nu$  &  $H_{ijk} u^i_\mu u^{j\mu} u^k_\nu \,\overline \psi \gamma^\nu \psi$ \\[-13pt]
    $\dim \ge 9$ &  &  \\[-13pt]

    & $G_{ijk} u^i_\mu u^j_\nu u^k_\alpha \epsilon^{\mu \nu \alpha \beta} j^\beta_\text{SM}$  &  $G_{ijk} u^i_\mu u^j_\nu u^k_\alpha \epsilon^{\mu \nu \alpha \beta} \, \overline \psi \gamma^\beta \psi$ \\ \hline

    \end{tabular}
    \caption{Operators coupling pGBs to the SM, organized by EFT dimension. The $\dim\ge5$ row is forbidden for DALPs by {\bf [C1]}. The leading DALP operators $\mathcal{O}_{Ha}$, $\mathcal{O}_{Ba}$ sit in the $\dim\ge6$ row. 
    See text for the definition of the SM operators and invariant tensors. Coset indices $i,j,k$ can be converted between upper and lower indices using the invariant tensors $\delta^{ij}$ and $\delta_{ij}$.     }
    \label{tab:operators}
\end{table}

\begin{itemize}
\item $\dim \ge 6$ row: $O_\text{SM}^{(\mu \nu)}$ denotes a SM operator which is symmetric and traceless, while $O_\text{SM}^{[\mu \nu]}$ is antisymmetric. Furthermore, $S_{ij}$ and $A_{ij}$ denote generic $H$-invariant symmetric and antisymmetric tensors (the same notation applies to the rest of the rows).  This row contains the LO DALP operators, since there is always a symmetric tensor $S_{ij}$, the Kronecker delta $\delta_{ij}$ in  e.g.\ the GB kinetic energy term.  If there is  an antisymmetric invariant, so that condition {\bf [C2]} is satisfied, then we also get a $B_{\mu \nu}$ operator starting at $d=6$, and so on. In particular, the two LO DALP operators already anticipated in \cref{eq:DALP-LO-Lag1,eq:DALP-LO-Lag,2.4} correspond to the two leading ones shown in this row,
\begin{equation}
S_{ij}\,u^{i\mu}u^j_\mu\,(H^\dagger H)
=
S_{ij}\,
\partial^\mu\theta^i\,
\partial_\mu\theta^j\,
(H^\dagger H)
+\mathcal O(\theta^3)\,,
\end{equation}
\begin{equation}
A_{ij}\,u^{i\mu}u^{j\nu}B_{\mu\nu}
=
A_{ij}\,
\partial^\mu\theta^i\,
\partial^\nu\theta^j\,
B_{\mu\nu}
+\mathcal O(\theta^3)\,.
\end{equation}
Furthermore, there are higher dimension  $d_\text{SM} \ge 4$ SM operators, such as any $d_\text{SM}=4$ term in the SM Lagrangian, $(D_\mu H)^\dagger (D_\nu H)$, etc. which give rise to other operators in the $\dim \ge 6$ row  with $\dim \ge 8$. 

\item $\dim \ge 7$ row: note that the dimension-six piece of the combination  $A_{ij}  u_i^\mu u_j^\nu O_\text{SM}^{[\mu \nu]}$ is absent as long as $\partial_\mu O_\text{SM}^{[\mu \nu]}=0$, e.g. the $d=6$ piece of  $A_{ij}  u_i^\mu u_j^\nu \widetilde B^{\mu \nu}$  has the form 
\begin{align}
\frac{A_{ij}}{f_{a,i} f_{a,j}} \partial^\mu a_i\, \partial^\nu a_j\, \widetilde B_{\mu \nu} & \to - \frac{A_{ij}}{f_{a,i} f_{a,j}} a_i\, \partial^\nu a_j\,  \partial^\mu \widetilde B_{\mu \nu}
\to 0
\,,
\label{3.23}
\end{align}
which vanishes by integration by parts and using the Bianchi identity $\partial^\mu   \widetilde B_{\mu \nu}=0$. Higher order terms in the $a/f_a$ expansion cannot be eliminated this way, since $u^\mu$ is not a total derivative.  Higher powers of $a/f$ contribute to the  $\widetilde B_{\mu \nu}$ operator starting at $d=7$ for non-symmetric cosets, and  at $d=8$ for symmetric cosets.

\item $\dim\ge 9$: $H_{ijk}$ denotes a generic invariant tensor which is completely symmetric or has mixed symmetry, while $G_{ijk}$ is a completely antisymmetric invariant tensor. For a semisimple group, one completely antisymmetric three-index tensor is the structure constants $f_{ijk}$. 

\item Each  operator in \cref{tab:operators} leads to a series of terms of increasing dimension from the expansion of $u_\mu^i$ in GB fields, \cref{1.9a}, with $\theta = a/f_a$. For a dark symmetry  breaking pattern $U(1)\to\varnothing$, $u_\mu^i = \partial_\mu \theta^i$ with no higher order terms. The expansion of $u_\mu^i$ has only odd powers of $\theta$ for symmetric cosets, and all powers of $\theta$ for non-symmetric cosets.

\end{itemize}

Note that \cref{eq:DALP-LO-Lag1} uses only the $d=6$ part of the chiral invariant form \cref{1.20} expanded in $a/f$.  Note as well that there are no higher order terms for $U(1)$ (flat) cosets.

\subsection{Implications of the conditions}\label{subsec:conditions}

We now explore the consequences on the dark sector of the two requirements in \cref{sec:chpt}, {\bf [C1]:}   $\RGB\not\supset \I_H$, so that the $d=5$ effective operators are not allowed, and  {\bf [C2]:} $(\RGB \otimes \RGB)_A \supset \I_H$, so that the distinctive operator $\mathcal{O}_{Ba}$ is allowed.

Condition {\bf [C1]} simply requires that $\RGB$ contains no $H$-invariant singlet: the  Higgs-DALP coupling $\mathcal{O}_{Ha}$  is among the LO contributions. Condition {\bf [C2]}, which allows the $B_{\mu \nu}$ operator $\mathcal{O}_{Ba}$,  requires more care. Any irreducible representation $\mathbf{r}$ of a given group $H$ is either \emph{real} ($\mathbf{r}\cong \bar{\mathbf{r}}$ with a symmetric bilinear invariant, $(\mathbf{r} \otimes \mathbf{r})_S\supset \I$), \emph{pseudo-real} ($\mathbf{p} \cong \bar{\mathbf{p}}$ with an antisymmetric bilinear invariant, $(\mathbf{p} \otimes \mathbf{p})_A\supset \I$), or \emph{complex} ($\mathbf{c}\not\cong \bar{\mathbf{c}}$).\footnote{The type of a given irrep can be determined by computing the Frobenius-Schur indicator, which takes the values $+1$, $0$, $-1$ for real, complex, and pseudo-real irreps, respectively.} The representation matrices for a real representation can be chosen to be real, but this is not possible for pseudoreal or complex representations. The adjoint representation of $G$ is real. Thus, under the decomposition \cref{1.1}, $\RGB$ contains a sum of
\begin{enumerate}
\item[(i)] real representations $\RGB\supset \mathbf{r}$ with any multiplicity,
\item[(ii)] pairs of identical pseudo-real irreps, $\RGB\supset \mathbf{p}\oplus \mathbf{p}$, so that pseudoreal irreps occur with even multiplicity,
\item[(iii)] pairs of complex-conjugate irreps, $\RGB\supset \mathbf{c}\oplus \mathbf{\overline c}$.
\end{enumerate}
In cases (ii) and (iii), there is an antisymmetric invariant. In case (i) with a single copy of $\mathbf{r}$, the invariant representation is in the symmetric product $(\mathbf{r} \otimes \mathbf{r})_S$, but not in the antisymmetric product. To have an antisymmetric invariant in case (i) requires more than one copy of $\mathbf{r}$ in $\RGB$. Thus to have an antisymmetric invariant, we need $\RGB$ to contain more than one copy of a real representation $\mathbf{r}$, or to contain a pair of pseudoreal representations, or to contain a complex representation and its conjugate.

 \subsubsection*{Refining the conditions} \label{Subsec:refining2}
 
 The  conditions above require explicit knowledge of the branching rule $\adj(G)\bigr|_H= \adj(H) \oplus \RGB$, or, in other words, knowledge of the representation $\RGB$ of the Goldstone bosons under $H$. We now present two alternative conditions that depend only on properties of the groups $G$ and $H$, without determining the GB representation $\RGB$ explicitly.

\paragraph*{\quad Equivalent condition for {\bf{[C1]}}.}
Condition {\bf [C1]} holds if and only if every $U(1)$ subgroup of $G$ that commutes with all of $H$ is contained in $H$, i.e. $G$ does not have a subgroup of the form $H \times U(1)$.

This equivalence is proved in App.~\ref{app:proofC1}. As a corollary, rank-preserving breakings ($\operatorname{rank}G=\operatorname{rank}H$) always satisfy {\bf [C1]}, though the converse does not hold in general. (QCD-like breaking patterns are a counterexample, see \cref{sec:examples}).

\paragraph*{\quad Sufficient condition for {\bf{[C2]}}.}
If $H$ has a $U(1)$ direct factor, $H\simeq H^\prime \times U(1)$, which is not a direct factor of $G$, then {\bf [C2]} is satisfied.

Note that this is only a sufficient (and not necessary) condition for {\bf [C2]}, but it is powerful enough to diagnose several of the examples discussed in \cref{sec:examples}. This condition can be further refined, as we prove in \cref{app:proofs}.

\subsection{DALP masses}\label{subsec:dalp_masses}

In the chiral limit, where $G$ is only spontaneously and not explicitly broken, all DALPs are exactly massless as true Goldstone bosons. It is however interesting to consider the case in which a small explicit breaking of $G$ gives them a mass $m_a \ll f_a$, so that they become pGBs. One may then ask what the DALP spectrum looks like under a small $G$-breaking that preserves $H$. For example, in QCD, a quark mass matrix proportional to the identity breaks chiral $SU(3) \times SU(3)$, but is invariant under vector $SU(3)$.

A mass term is a symmetric bilinear in the DALP fields, so it must be an $H$-invariant in the symmetric product $(\RGB \otimes \RGB)_S$,  since $H$ stays exact. All DALPs within an irreducible multiplet are related by the symmetry and must be degenerate, much as isospin makes the pions degenerate. That is, DALPs in the same $H$-irrep share a common mass, while different irreps may get different masses. Furthermore, for a complex irrep, the whole  conjugate pair $\mathbf{c}\oplus\bar{\mathbf{c}}$ is degenerate.  As an example, in the symmetric coset $SU(2)\to U(1)$, the two DALPs carry opposite $U(1)$ charge and form a  complex conjugate pair built from two real fields, so $m_{a_1}=m_{a_2}$. For $SU(3)\to U(1)\times U(1)$, the broken generators split into three such charged pairs, which generically acquire three different masses, each pair consisting of two degenerate real fields (see decomposition in \cref{2.3}).

\subsection{Examples}\label{sec:examples}

It is useful to consider some examples. First, we consider two simple cases which do not satisfy conditions  {\bf{[C1, C2]}}, which are the cases extensively studied in the ALP literature.
\begin{itemize}
\item Any $U(1)$ breaking, such as $U(1) \to \varnothing$ or $U(1) \times U(1) \to \varnothing$, has GBs which are invariant under $H$, since the unbroken group $H$ is trivial and consists only of the identity. This example does not satisfy condition  {\bf{[C1]}}.
\item QCD-like breaking $SU(N)_L \times SU(N)_R \to SU(N)_V$. In this case, the GBs transform as $\RGB=\adj \, SU(N)$, so {\bf{[C1]}} is satisfied. However, the adjoint representation is real, so the symmetric product $(\RGB \otimes \RGB)_S$ contains an $H$-invariant, but the antisymmetric product $(\RGB \otimes \RGB)_A$ does not, so {\bf{[C2]}} is not satisfied. The first interaction of DALPs with the SM arises at $d=6$ with $ S_{ij} u_i^\mu u_j^\mu H^\dagger H$ followed by the $d=8$ interaction $ S_{ij} u_i^\mu u_j^\mu O_\text{SM}$ where $O_\text{SM}$ is a dimension-four term in the SM Lagrangian.\footnote{{This only holds in the chiral limit, i.e. whenever $G$ is only spontaneously and not explicitly broken, and if the SM fields are not charged under $G$. Both assumptions are violated in QCD, and thus an analogous operator is generated, $u^\mu_i u^\nu_j f_{ija} F^a_{\mu\nu}$, see e.g.\ Eq.~(4.159) in Ref.~\cite{Scherer:2002tk}.} }
\end{itemize}

We next give two examples which do meet both conditions, and thus allow for both the Higgs-portal and  $B_{\mu\nu}$ operator in the $d=6$ interaction Lagrangian: (a) $SU(2) \to U(1)$, which is equivalent to $SO(3) \to SO(2)$, and (b) $SU(3) \to U(1) \times U(1)$.  In both examples, $G$ is broken to its Cartan subgroup.  Example (a) illustrates the case of a symmetric coset, while example (b) has a non-symmetric coset.

\subsubsection{Example (a): Symmetric Coset $SU(2)\to U(1)$} \label{subsubsec:Exa}

The breaking $SU(2) \to U(1)$ where the $U(1)$ generator is $T_3$  has
\begin{align}
\adj\, G  &= \mathbf{3},  & \adj\, H &= (0), &   \RGB &= (1) \oplus (-1) \, .
\label{2.1}
\end{align}
The CCWZ form \cref{1.2} is
\begin{align}
\bm{\theta} &= \frac12 \begin{bmatrix} 0 & \theta_1 - i \theta_2 \\  \theta_1 + i \theta_2 & 0 \end{bmatrix} \, ,
\label{2.2}
\end{align}
using the standard labelling of $SU(2)$ matrices, and using real fields instead of complex charged fields. The fields $(a_1\mp i a_2)/\sqrt{2}$ have charge $\pm1$. The expansions of $u_i^\mu$, $i=1,2$,
\begin{align}
u_1^\mu &=  \partial^\mu \theta_1 + \frac16 \theta_2 \left(\theta_1 \partial^\mu \theta_2 - \theta_2 \partial^\mu \theta_1 \right)  + \ldots \nn \\
u_2^\mu & = \partial^\mu \theta_2 - \frac16 \theta_1 \left(\theta_1 \partial^\mu \theta_2 - \theta_2 \partial^\mu \theta_1 \right)  + \ldots \, ,
\label{2.15}
\end{align}
have only odd powers of $\theta$, since the coset is symmetric. This example is simple enough that the expansion can be computed to all orders (see e.g.\ Ref.~\cite[Appendix~A]{Alonso:2016oah}). There is a unique two-index symmetric invariant $\delta_{ij}$ and a unique two-index antisymmetric invariant $\epsilon_{ij}$, where $\epsilon_{12}=1$. The kinetic term is
\begin{align}
\mathcal{L} &=\frac12 f_a^2 \delta_{ij} u_i^\mu u_j^\mu = \frac12 \partial_\mu a_i \partial^\mu a_i -\frac1{6f_a^2} (a_1 \partial_\mu a_2-a_2 \partial_\mu a_1)^2+ \ldots
\label{2.6}
\end{align}
where $\theta_i=a_i/f_a$. The antisymmetric invariant which multiplies $B_{\mu \nu}$ or $\widetilde B_{\mu \nu}$ in \cref{1.20} is
\begin{align}
\epsilon_{ij} u_i^\mu u_j^\nu &=\frac{1}{f_a^2} \left( \partial^\mu a_1 \partial^\nu a_2 -  \partial^\mu a_2 \partial^\nu a_1 \right) \left[ 1 - \frac{1}{6f_a^2}
( (a_1)^2 + (a_2)^2) + \ldots \right].
\label{2.7}
\end{align}

\subsubsection{Example (b): Non-symmetric Coset $SU(3)\to U(1) \times U(1)$}  \label{subsubsec:Exb}

The breaking $SU(3) \to U(1) \times U(1) $, where the $U(1)$ generators are normalized to $T_3$ and $\sqrt{12} \, T_8$, has
\begin{align}
\adj\, G  &= \mathbf{8},  \nn \\
\adj\, H &= (0,0) \oplus (0,0), \nn \\
\RGB &= (1,0) \oplus (-1,0) \oplus ({\scriptstyle \frac12} ,3) \oplus({\scriptstyle \frac12},-3) \oplus (-{\scriptstyle \frac12},-3) \oplus (-{\scriptstyle \frac12},3) \ .
\label{2.3}
\end{align}
The CCWZ form \cref{1.2} is
\begin{align}
\bm{\theta} &= \frac12 \begin{bmatrix} 0 & \theta_1 - i \theta_2 & \theta_4 - i \theta_5  \\  \theta_1 + i \theta_2 & 0 & \theta_6 - i \theta_7 \\
 \theta_4 + i \theta_5 & \theta_6 + i \theta_7 & 0 \end{bmatrix} \, ,
\label{2.14}
\end{align}
using the standard labeling of $SU(3)$ matrices, and using real fields instead of complex charged fields. The fields $(a_1\mp i a_2)/\sqrt{2} $ have charge $\pm (1,0)$, $(a_4\mp i a_5)/\sqrt{2}$ have charge $\pm ({\scriptstyle \frac12},3)$ and $(a_6 \mp i a_7)/\sqrt{2}$ have charge $\pm (-{\scriptstyle \frac12},3)$. The fields $u_i^\mu$ have an expansion that has all powers of $\theta$, since the coset is not symmetric. For example,
\begin{align}
u_1^\mu &= \partial^\mu \theta_1 
+ \frac14 ( \theta_4\, \partial^\mu \theta_7 - \theta_7 \, \partial^\mu \theta_4)  + \frac14 (\theta_6\, \partial^\mu \theta_5 - \theta_5\, \partial^\mu \theta_6 )+ O\left( \theta^3\right) \,.
\label{2.8}
\end{align}

The Goldstone boson representation $\RGB$ is reducible, and there are three symmetric invariants and three antisymmetric invariants: $\delta_{ij}$ and $\epsilon_{ij}$, in the three subspaces $ij=12$, $45$, and $67$, which we denote by $\delta^{(12)}_{ij}$, $\epsilon^{(12)}_{ij}$, etc. The most general kinetic term is
\begin{align}
\mathcal{L} &= \frac12 \left( f_{12}^2\, \delta^{(12)}_{ij} + f_{45}^2\, \delta^{(45)}_{ij} + f_{67}^2\, \delta^{(67)}_{ij} \right) u^i_\mu u^j_\mu \, ,\nn \\
&= \frac12 \partial_\mu a_i \partial_\mu a_i + \mathcal{O}\left( \frac{1}{f_a} a\, \partial_\mu a\, \partial_\mu a \right)\,,
\label{2.9}
\end{align}
where for a properly normalized kinetic term, $\theta_i=a_i/f_i$, and $f_i=f_{12}$ for $i=1,2$, $f_i=f_{45}$ for $i=4,5,$ and $f_i=f_{67}$ for $i=6,7$. Notice that different Goldstone bosons can have different decay constants. The antisymmetric invariants are
\begin{align}
\epsilon^{(12)}_{ij} u^i_\mu u^j_\nu &= \frac{1}{f_{12}^2} \left( \partial_\mu a_1\, \partial_\nu a_2 - \partial_\mu a_2 \, \partial_\nu a_1 \right)
+\frac1{4 f_{12} f_{45}f_{67}} a_4 ( \partial_\mu a_1\,  \partial_\nu a_6 -  \partial_\mu a_6\, \partial_\nu a_1     ) +  \ldots  \nn \\
\epsilon^{(45)}_{ij} u^i_\mu u^j_\nu &= \frac{1}{f_{45}^2} \left( \partial_\mu a_4\, \partial_\nu a_5 - \partial_\mu a_5\, \partial_\nu a_4 \right) + \ldots  \nn \\
\epsilon^{(67)}_{ij} u^i_\mu u^j_\nu &= \frac{1}{f_{67}^2} \left( \partial_\mu a_6\,  \partial_\nu a_7 - \partial_\mu a_7\, \partial_\nu a_6 \right) + \ldots
\label{2.10}
\end{align}
with $\ldots$ of order $\mathcal{O}(a_i^3)$. We have given one sample $\mathcal{O}(a^3)$ term in the first expression.

To see how a kinetic term such as \cref{2.9} arises, consider the breaking $SU(3) \to U(1) \times U(1)$ by the vacuum expectation value of an $SU(3)$ adjoint field $\Phi$,
\begin{align}
\vev{\Phi} &= \begin{bmatrix} v_1 & 0 & 0 \\ 0 & v_2 & 0 \\ 0 & 0 & -v_1 -v_2 \end{bmatrix} \ .
\label{2.11}
\end{align}
The fluctuations of $\Phi$ in the Goldstone boson directions can be parameterized by
\begin{align}
\Phi &= \xi \vev{\Phi}  \xi^{-1} \, ,
\label{2.12}
\end{align}
and the $\Phi$ kinetic energy term gives
\begin{align}
\mathcal{L} &= \frac12 \tr \partial_\mu \Phi \partial_\mu \Phi =\frac14 \left( (v_1-v_2)^2\, \delta^{(12)}_{ij} + (2v_1+v_2)^2\, \delta^{(45)}_{ij} + (v_1+2v_2)^2\, \delta^{(67)}_{ij} \right) u^i_\mu u^j_\mu \, ,
\label{2.13}
\end{align}
so that in \cref{2.9}, $f_{12}^2= (v_1-v_2)^2/2$, $f_{45}^2= (2v_1+v_2)^2/2$, $f_{67}^2= (v_1+2v_2)^2/2$.  It is not possible to make all three decay constants equal for any choice of $v_1$ and $v_2$.
The most general symmetric tensor is
\begin{align}
S_{ij} &= c_{12}\,  \delta^{(12)}_{ij} + c_{45}\,  \delta^{(45)}_{ij} + c_{67}\, \delta^{(67)}_{ij} \, ,
\label{2.21}
\end{align}
and the linear combination of the three possible invariants for higher dimension invariants such as $S_{ij} u^i_\mu u^j_\mu (H^\dagger H)$ can be different from the kinetic term. Similarly, the most general antisymmetric invariant is a linear combination of  $\epsilon^{(12)}_{ij} $, $\epsilon^{(45)}_{ij} $ and $\epsilon^{(67)}_{ij} $.

\subsection{Other examples}\label{subsubsec:OtherSBBpatterns}

Some other symmetry breaking patterns which satisfy conditions {\bf [C1]} and {\bf [C2]} are:

\begin{itemize}

\item $SU(p+q) \to SU(p) \times SU(q) \times U(1)$. In this case, $\RGB = (\mathbf{p},\,\bar{\mathbf{q}})_1 \oplus (\bar{\mathbf{p}},\,\mathbf{q})_{-1}$ for a suitable normalization of the $U(1)$ generator.  $\RGB$ has no singlet, so {\bf [C1]}  holds, and the pairing between $(\mathbf{p},\bar{\mathbf{q}})_1$ and $(\bar{\mathbf{p}},\mathbf{q})_{-1}$ gives an antisymmetric invariant, so {\bf [C2]} holds.  Familiar special cases are $SU(3) \to SU(2) \times U(1)$, with $\RGB= \mathbf{2}_1 \oplus \mathbf{2}_{-1}$ and $SU(5) \to SU(3) \times SU(2) \times U(1)$ with $\RGB = (\mathbf{3},\mathbf{2})_1 \oplus (\mathbf{\bar 3}, \mathbf{2})_{-1}$.

\item $SU(N) \to U(1)^{N-1}$, where $SU(N)$ breaks to the Cartan subalgebra. $\RGB$ consists of a collection of charges $(q_1,\cdots, q_{N-1})$ and their complex conjugates. There are no zero charge particles, so {\bf [C1]} holds, and each complex conjugate pair provides an antisymmetric invariant, so {\bf [C2]} holds.

\item Other families include
\begin{align}
  SO(2n) &\to U(n)\,, \nonumber \\
  Sp(2n) &\to U(n)\,, \nonumber \\
  SO(n{+}2) &\to SO(n)\times SO(2)\,.
\end{align}

\end{itemize}
It is easy to verify the refined conditions on these patterns. {\bf [C1]} follows from the breaking being rank-preserving and {\bf [C2]} from the fact that all the unbroken subgroups $H$ contain a $U(1)$ factor that is not a direct factor of $G$.

\section{Phenomenological constraints on the LO DALP Lagrangian}\label{sec:pheno}

\begin{figure}
\centering
\resizebox{\textwidth}{!}{%
\begin{tikzpicture}
\tikzset{photon/.style={decorate, decoration={snake, segment length=1.75mm, amplitude=0.75mm}}}
%
\begin{scope}[xshift=0cm]
\coordinate (v) at (0,0);
\draw[dashed] (v) -- +(180:1.3);
\draw[dashed] (v) -- +(30:1.2);
\draw[dashed] (v) -- +(-30:1.2);
\filldraw (v) circle (0.07);
\draw (v)+(180:1.6) node {$h$};
\draw (v)+(30:1.5) node {$a_i$};
\draw (v)+(-30:1.5) node {$a_i$};
\draw (0,-1.5) node {(a)};
\end{scope}
\begin{scope}[xshift=4cm]
\coordinate (v) at (0,0);
\draw[dashed] (v) -- +(135:1.2);
\draw[dashed] (v) -- +(225:1.2);
\draw[dashed] (v) -- +(45:1.2);
\draw[dashed] (v) -- +(-45:1.2);
\filldraw (v) circle (0.07);
\draw (v)+(135:1.55) node {$h$};
\draw (v)+(225:1.55) node {$h$};
\draw (v)+(45:1.55) node {$a_i$};
\draw (v)+(-45:1.55) node {$a_i$};
\draw (0,-1.5) node {(b)};
\end{scope}
\begin{scope}[xshift=8cm]
\coordinate (v) at (0,0);
\draw[photon] (v) -- +(180:1.3);
\draw[dashed] (v) -- +(30:1.2);
\draw[dashed] (v) -- +(-30:1.2);
\filldraw (v) circle (0.07);
\draw (v)+(180:1.6) node {$\gamma$};
\draw (v)+(30:1.5) node {$a_1$};
\draw (v)+(-30:1.5) node {$a_2$};
\draw (0,-1.5) node {(c)};
\end{scope}
\begin{scope}[xshift=12cm]
\coordinate (v) at (0,0);
\draw[photon] (v) -- +(180:1.3);
\draw[dashed] (v) -- +(30:1.2);
\draw[dashed] (v) -- +(-30:1.2);
\filldraw (v) circle (0.07);
\draw (v)+(180:1.6) node {$Z$};
\draw (v)+(30:1.5) node {$a_1$};
\draw (v)+(-30:1.5) node {$a_2$};
\draw (0,-1.5) node {(d)};
\end{scope}
\end{tikzpicture}%
}
\caption{\label{fig:brokenvertices} Effective DALP vertices in the broken electroweak phase. $\mathcal{O}_{Ha}$ generates (a) $h\, a_i a_i$ and (b) $h h\, a_i a_i$; $\mathcal{O}_{Ba}$ generates (c) $\gamma\, a_1 a_2$ and (d) $Z\, a_1 a_2$.}
\end{figure}
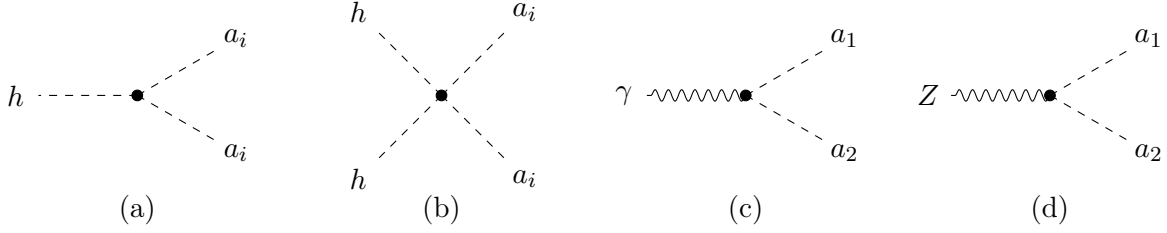

The phenomenological signals of DALPs  differ drastically from those for the customary uncharged ALPs as all  $d=5$ effective couplings are forbidden. In particular, the paradigmatic ALP-two photon coupling, which provides strong bounds for uncharged ALPs, does not exist. Novel signals are expected instead for DALPs, among which are Higgs-two DALP, two Higgs-two DALP, photon-two DALP, and $Z$-two DALP   interactions (see  \cref{fig:LOoperators}  and  \cref{fig:brokenvertices}), as we expatiate next.

\begin{figure}
    \centering
 
\hspace*{-0.75cm} \includegraphics[width=1.1\textwidth]{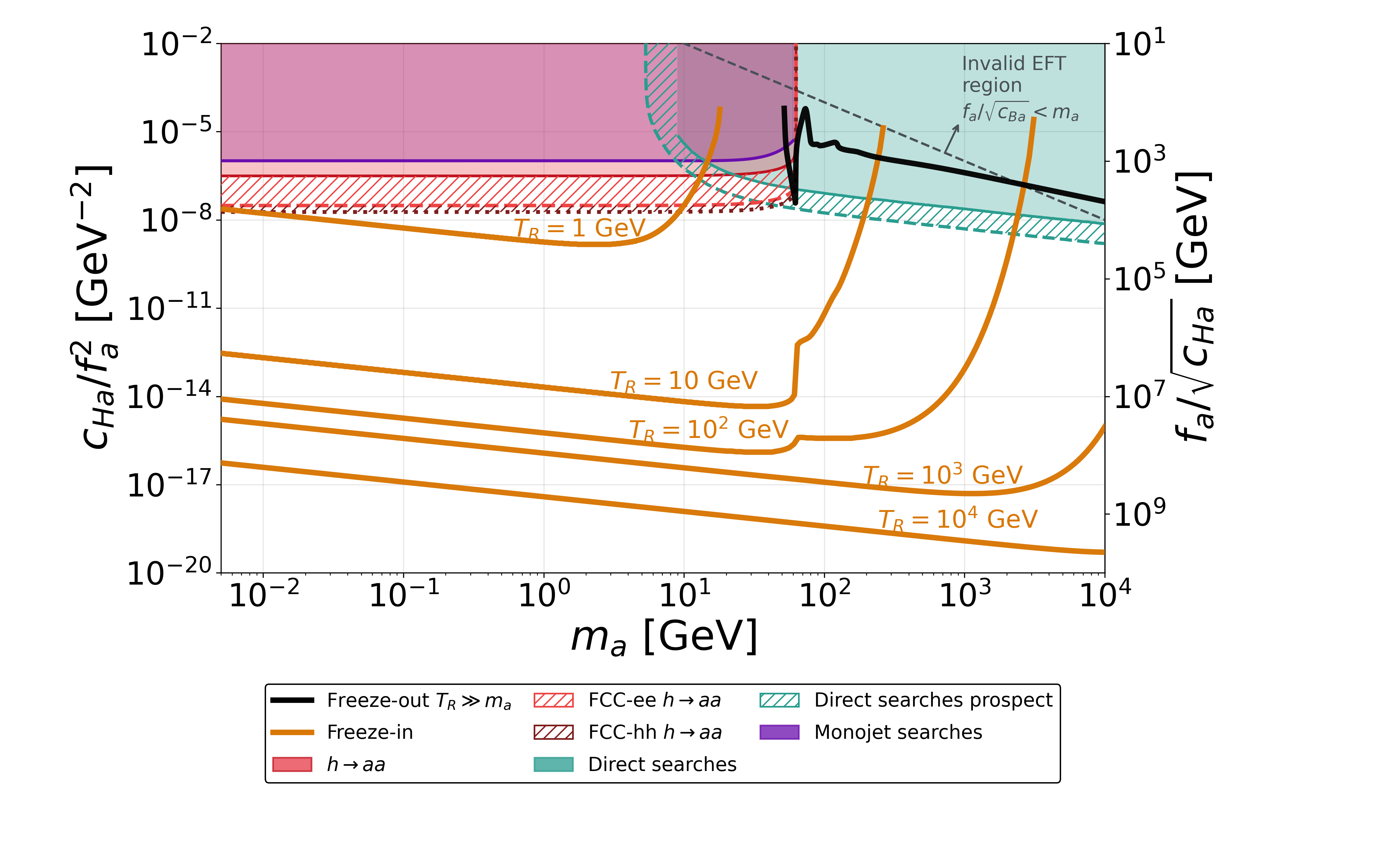}

    \caption{\small Constraints on the operator coefficients in the $(m_a,\, c_{Ha}/f_a^2)$ plane. The solid  black line corresponds to the correct DM relic abundance through freeze-out production for sufficiently high $T_R$, while orange lines are for freeze-in production for some example values of $T_R$. The shaded regions indicate the areas excluded by the Higgs invisible width (red), direct detection limits (green) and monojet searches (purple). Hatched regions are prospects for the invisible widths constrained by the FCC and direct searches from XLZD. The EFT breaks down in  and above the gray oblique discontinuous line.}
     \label{fig:EstudioH2}

\end{figure}

\begin{figure}
    \centering

  \hspace*{-0.75cm}  \includegraphics[width=1.1\textwidth]{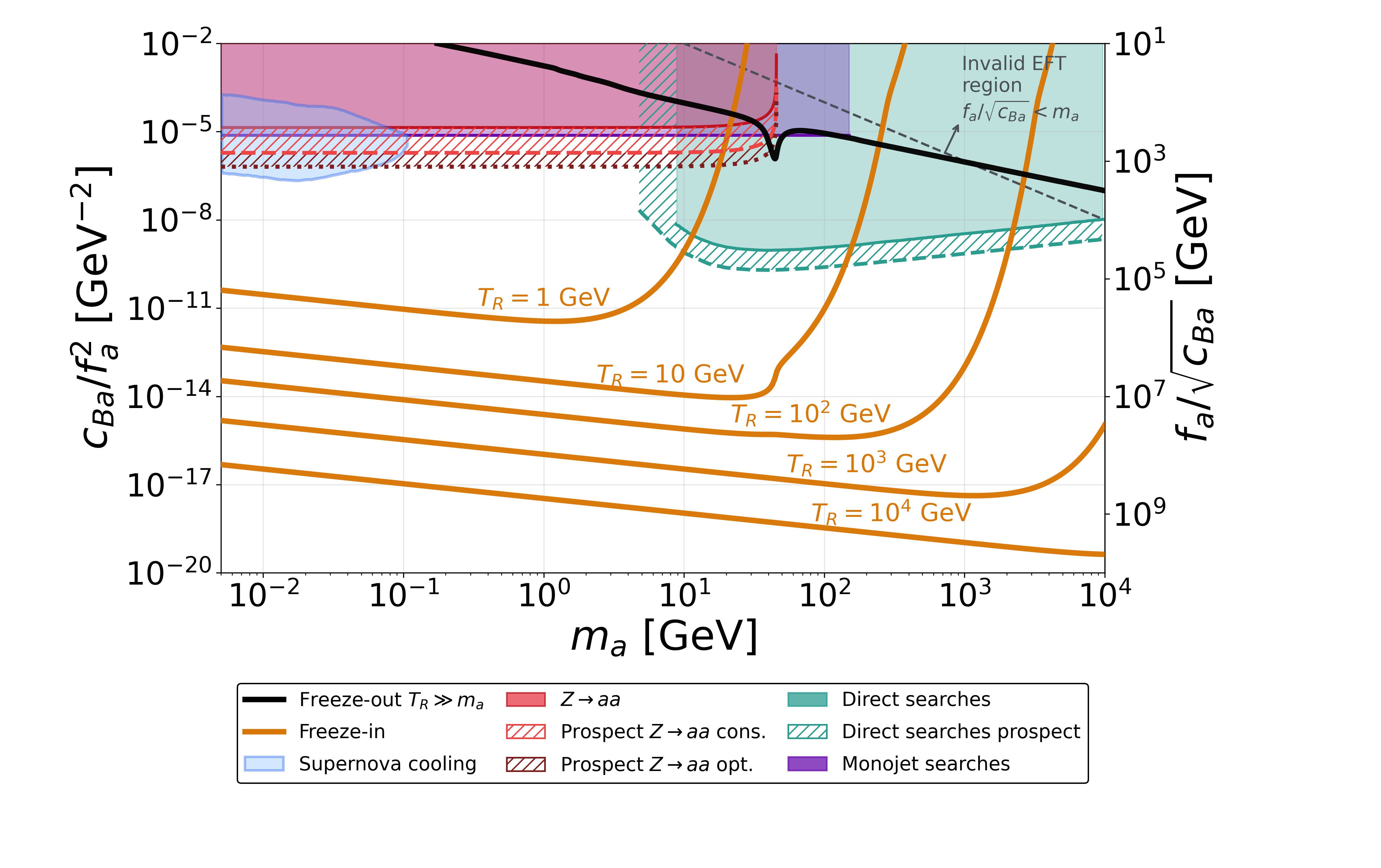}

    \caption{\small Constraints on the operator coefficients  in the  $(m_a,\, c_{Ba}/f_a^2)$ plane. The solid  black line corresponds to the correct DM relic abundance through freeze-out production for sufficiently high $T_R$, while orange lines are for freeze-in production for some example values of $T_R$. The shaded regions indicate the areas excluded by the $Z$ invisible width (red), direct detection limits (green), monojet searches (purple) and supernovae cooling (blue). Hatched regions are prospects for the invisible widths constrained by the FCC and direct searches from XLZD. The EFT breaks down in  and above the gray oblique discontinuous line.}
    \label{fig:EstudioZ2}
    
\end{figure}

DALPs can be pair-produced, and they can be probed through invisible decays as well as monojet or monophoton searches at colliders. Moreover, the lightest DALP is stable and a potential dark matter candidate. Below we briefly discuss each constraint derived from these observables and estimate their magnitude.  We summarize all constraints, as well as the parameter space where the correct DM abundance is obtained, in ~\cref{fig:EstudioH2,fig:EstudioZ2,fig:misalignment}. For the phenomenological constraints, we focus on the minimal case of {\it only two} DALPs, $\{a_1,a_2\}$, assumed to be {\it degenerate}, to avoid additional parameters. DALPs of this type occur in $G/H$ models where a small $H$-invariant, but $G$ non-invariant, term (whose form does not need to be specified) is added to the Lagrangian. This is analogous to how, in QCD, a small quark mass term proportional to the unit matrix leads to degenerate light pions.

\subsection{Invisible widths}\label{sec:invwidth}

The LO operators in \cref{eq:DALP-LO-Lag} lead to a pair of DALPs coupled to the Higgs boson ($\mathcal{O}_{Ha}$), and  to the hypercharge gauge boson ($\mathcal{O}_{Ba}$) which induces both photon and $Z$ couplings. The most direct and model-independent way to probe for them is through their contribution to the invisible widths of the Higgs boson and the $Z$. 

Considering two degenerate DALP species, we obtain the  width for the Higgs boson decay to DALP pairs,
\begin{equation}
   \sum_{i=1,2} \Gamma(h \to a_i a_i)
=
\frac{v^2c_{Ha}^2}{16\pi\, m_h\, Z_a^2f_a^4}
\,(m_h^2 - 2 m_{a}^2)^2\,
\sqrt{1 - \frac{4 m_{a}^2}{m_h^2}} \, ,
\label{hdecay}
\end{equation}
where $ Z_a\equiv 1+{c_{Ha} v^2}/{f_a^2}$  is the factor obtained after the field rescaling required to canonically normalize the DALP kinetic term after electroweak symmetry breaking, and $m_a$ denotes the DALP mass. The present $95\%$ C.L. bound on the invisible branching ratio of the Higgs boson~\cite{ParticleDataGroup:2024cfk} combined with the SM Higgs width of  $4.07\,\text{MeV}$ gives
\begin{equation}
    \Gamma(h \to {inv}) <  \frac{\text{BR}(h \to inv)}{1-\text{BR}(h \to inv)} \Gamma_\text{SM} = 0.49\,\mathrm{MeV}\, .
    \label{4.2}
\end{equation}

A slightly stronger constraint is obtained indirectly from constraints on the Higgs signal strength. Introducing a new invisible channel for Higgs boson decay reduces all the branching ratios for Higgs  decay and, consequently, the inclusive Higgs signal strength  is $\mu = 1-\Gamma(h \to a a)/\Gamma_\text{tot}$. 
Since both CMS~\cite{CMS:2026nce}, $\mu_\text{CMS}=1.014 \genfrac{}{}{0pt}{}{+0.055} {- 0.053}$, and ATLAS~\cite{ATLAS:2022vkf}, $\mu_\text{ATLAS}=1.05\pm0.06$, report a best-fit value for the inclusive signal strength $\mu$ slightly above the SM prediction, which is in the unphysical region since the presence of DALPs can only decrease the branching ratios for the visible channels, we combine their results following the prescription in Ref.~\cite{Feldman:1997qc} to obtain: 
\begin{equation}
    \Gamma(h \to inv) < 0.22\,\mathrm{MeV}\,,
\end{equation}
at the $95\%$ C.L. We use this constraint to derive a bound on the first operator of \cref{eq:DALP-LO-Lag} displayed in \cref{fig:EstudioH2} by the red area. The hatched red area depicts the constraint that could be derived from the expected sensitivity of the HL-LHC, FCC-ee and FCC-hh to the invisible Higgs decay. In~\cref{tab:CH-inv} we compile the present bounds and expected sensitivities from HL-LHC and FCC to the invisible Higgs decay as well as their corresponding constraints on $c_{Ha}$ in the $m_a=0$ limit.

\begin{table}
    \centering
    \begin{tabular}{|l|c|c|c|c}
    \hline
         &   BR$(h \to inv)$ & $\Gamma(h \to inv)$ & $c_{Ha} / f_a^2$ \\
        \hline
        Present observations & $5\%$ &  0.22\,MeV&  $3.1 \times 10^{-7}\,\mathrm{GeV}^{-2}$ \\
        \hline
        HL-LHC 
         & $1.9\%$ & 79\,keV & $1.8 \times 10^{-7}\,\mathrm{GeV}^{-2}$\\
        \hline
        FCC-ee
         & $0.055\%$ & 2.2\,keV & $3.1\times  10^{-8}\,\mathrm{GeV}^{-2}$ \\
        \hline
        FCC-hh  & $0.02\%$ & 0.81\,keV & $  1.9\times  10^{-8}\,\mathrm{GeV}^{-2} $ \\
        \hline
    \end{tabular}
\caption{Present bounds and future sensitivies to the branching ratio of $h \to inv$ and the corresponding bound on $c_{Ha}$ for $m_a=0$.}
    \label{tab:CH-inv}
\end{table}

The decay width of the $Z$ boson to DALPs is given by 
\begin{equation}
\Gamma(Z \to a_1 a_2) = \frac{\sin^2 \theta_Wc_{Ba}^2}{48 \pi} \frac{M_Z^5}{f_a^4} \left(1 - \frac{4m_a^2}{M_Z^2}\right)^{3/2} \, .
\label{eq:Z_decay_width}
\end{equation}
The final determination of the invisible $Z$ width by LEP~\cite{ALEPH:2005ab,Janot:2019oyi} is usually expressed as a measurement of the number of neutrinos $N_\nu=2.9963\pm0.0074$. Following again the prescription in Ref.~\cite{Feldman:1997qc} to infer how large the decay width $\Gamma(Z \to a_i a_j)$ is allowed to be, we find
\begin{equation}
    \Gamma(Z \to a_i a_j) < 1.84 \,\mathrm{MeV}\,,
\end{equation}
at the $95\%$~C.L.\ This bound corresponds to the excluded red area in \cref{fig:EstudioZ2}. The FCC-ee is expected to improve the precision on the measurement of the $Z$ width to 4~keV~\cite{deBlas:2025gyz}. Taken at face value, this would correspond to the exclusion region depicted by the hatched brown area. Nevertheless, it is not clear if the SM prediction will be able to match this sensitivity. A conservative estimate of the SM uncertainty of 35~keV was presented in Ref.~\cite{deBlas:2025gyz}, and would correspond to the hatched red exclusion region. It follows  from \cref{eq:Z_decay_width} that, for negligible DALP masses, the present, conservative FCC-foreseen, and optimistic FCC-foreseen sensitivities translate into $c_{Ba}/f_a^2 <  1.4\times 10^{-5}\,\mathrm{GeV}^{-2}$, $c_{Ba}/f_a^2 <  1.9\times 10^{-6}\,\mathrm{GeV}^{-2}$ and $c_{Ba}/f_a^2 <  6.4\times 10^{-7}\,\mathrm{GeV}^{-2}$,  respectively (see \cref{fig:EstudioZ2}). 

\subsection{Monojet searches}

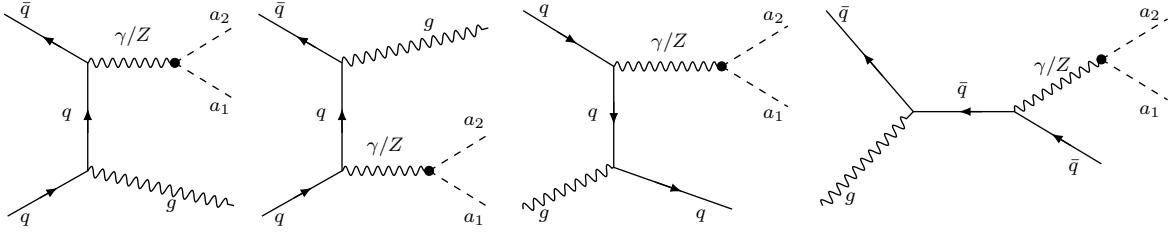
\begin{figure}
\centering
\resizebox{\textwidth}{!}{%
\begin{tikzpicture}[every node/.style={font=\scriptsize}]

\tikzset{
    photon/.style={
        decorate,
        decoration={snake, segment length=1.55mm, amplitude=0.65mm}
    },
    gluon/.style={
        decorate,
        decoration={coil, aspect=0, segment length=1.65mm, amplitude=0.75mm}
    },
    scalar/.style={dashed, line width=0.55pt},
    fermion/.style={line width=0.55pt},
    arrow/.style={-{Latex[length=1.7mm]}, line width=0.55pt}
}

\def\linearrow#1#2#3#4{%
    \draw[arrow] ($(#1)!#3!(#2)$) -- ($(#1)!#4!(#2)$);
}

\begin{scope}[xshift=0cm]

\coordinate (qin) at (-1.15,-0.65);
\coordinate (v1) at (0,0);
\coordinate (v2) at (0,1.55);
\coordinate (qbarin) at (-1.20,2.25);
\coordinate (vd) at (1.25,1.55);

\draw[fermion] (qin) -- (v1) -- (v2) -- (qbarin);

\linearrow{qin}{v1}{0.40}{0.63}
\linearrow{v1}{v2}{0.35}{0.58}
\linearrow{v2}{qbarin}{0.32}{0.55}

\draw[photon, line width=0.55pt] (v2) -- (vd);
\node at (0.65,1.93) {$\gamma/Z$};

\filldraw (vd) circle (0.065);

\draw[scalar] (vd) -- +(0.90,0.55);
\draw[scalar] (vd) -- +(0.90,-0.55);

\node at (1.9,2.20) {$a_2$};
\node at (1.9,0.90) {$a_1$};

\draw[gluon, line width=0.55pt] (v1) -- +(2.10,-0.50);
\node at (1.2,-0.5) {$g$};

\node at (-0.9,-0.70) {$q$};
\node at (-0.3,0.80) {$q$};
\node at (-0.9,2.30) {$\bar q$};

\end{scope}

\begin{scope}[xshift=3.65cm]

\coordinate (qin) at (-1.15,-0.65);
\coordinate (v1) at (0,0);
\coordinate (v2) at (0,1.55);
\coordinate (qbarin) at (-1.20,2.25);
\coordinate (vd) at (1.25,0);

\draw[fermion] (qin) -- (v1) -- (v2) -- (qbarin);

\linearrow{qin}{v1}{0.40}{0.63}
\linearrow{v1}{v2}{0.35}{0.58}
\linearrow{v2}{qbarin}{0.32}{0.55}

\draw[gluon, line width=0.55pt] (v2) -- +(2.10,0.50);
\node at (1.25,2.1) {$g$};

\draw[photon, line width=0.55pt] (v1) -- (vd);
\node at (0.62,0.35) {$\gamma/Z$};

\filldraw (vd) circle (0.065);

\draw[scalar] (vd) -- +(0.90,0.55);
\draw[scalar] (vd) -- +(0.90,-0.55);

\node at (1.9,0.7) {$a_2$};
\node at (1.9,-0.70) {$a_1$};

\node at (-0.9,-0.70) {$q$};
\node at (-0.3,0.80) {$q$};
\node at (-0.9,2.30) {$\bar q$};

\end{scope}

\begin{scope}[xshift=7.55cm, yshift=-0.05cm]

\coordinate (qin) at (-1.30,2.35);
\coordinate (v1) at (0,1.55);
\coordinate (v2) at (0,0.10);
\coordinate (qout) at (1.70,-0.50);
\coordinate (vd) at (1.55,1.55);

\draw[fermion] (qin) -- (v1) -- (v2) -- (qout);

\linearrow{qin}{v1}{0.38}{0.60}
\linearrow{v1}{v2}{0.35}{0.58}
\linearrow{v2}{qout}{0.35}{0.58}

\draw[gluon, line width=0.55pt] (-1.30,-0.50) -- (v2);
\node at (-1.0,-0.6) {$g$};

\draw[photon, line width=0.55pt] (v1) -- (vd);
\node at (0.8,1.9) {$\gamma/Z$};

\filldraw (vd) circle (0.065);

\draw[scalar] (vd) -- +(0.95,0.58);
\draw[scalar] (vd) -- +(0.95,-0.58);

\node at (2.3,2.25) {$a_2$};
\node at (2.3,0.85) {$a_1$};

\node at (-1.00,2.35) {$q$};
\node at (-0.3,0.85) {$q$};
\node at (1.25,-0.6) {$q$};

\end{scope}

\begin{scope}[xshift=11.85cm, yshift=-0.05cm]

\coordinate (qbarin) at (-1.25,2.35);
\coordinate (v1) at (0,0.90);
\coordinate (v2) at (1.45,0.90);
\coordinate (qbarout) at (2.70,0.15);
\coordinate (vd) at (2.70,1.65);

\draw[fermion] (qbarin) -- (v1) -- (v2) -- (qbarout);

\linearrow{v1}{qbarin}{0.35}{0.58}
\linearrow{v2}{v1}{0.32}{0.55}
\linearrow{qbarout}{v2}{0.35}{0.58}

\draw[gluon, line width=0.55pt] (-1.30,-0.45) -- (v1);
\node at (-0.90,-0.35) {$g$};

\draw[photon, line width=0.55pt] (v2) -- (vd);
\node at (2.0,1.55) {$\gamma/Z$};

\filldraw (vd) circle (0.065);

\draw[scalar] (vd) -- +(0.95,0.58);
\draw[scalar] (vd) -- +(0.95,-0.58);

\node at (3.4,2.30) {$a_2$};
\node at (3.4,0.90) {$a_1$};

\node at (-1.0,2.3) {$\bar q$};
\node at (0.7,1.2) {$\bar q$};
\node at (2.3,0.1) {$\bar q$};

\end{scope}

\end{tikzpicture}
}

\caption{Dominant Feynman diagrams for the monojet processes with the insertion of the $O_{Ba}$ operator represented by a thick dot, and where $q=u,d$.}
\label{fig:monojet_diagrams_c_ba}
\end{figure}

\begin{figure}
\centering
\resizebox{0.78\textwidth}{!}{%
\begin{tikzpicture}[every node/.style={font=\scriptsize}]

\tikzset{
    gluon/.style={
        decorate,
        decoration={coil, aspect=0, segment length=1.65mm, amplitude=0.75mm}
    },
    scalar/.style={dashed, line width=0.55pt},
    fermion/.style={line width=0.55pt},
    arrow/.style={-{Latex[length=1.7mm]}, line width=0.55pt}
}

\def\linearrow#1#2#3#4{%
    \draw[arrow] ($(#1)!#3!(#2)$) -- ($(#1)!#4!(#2)$);
}

\begin{scope}[xshift=0cm]

\coordinate (vbot) at (0,0);
\coordinate (vtop) at (0,1.25);
\coordinate (vh)   at (1.25,0);

\coordinate (gin)  at (-1.05,-0.60);
\coordinate (qin)  at (-1.10,2.00);
\coordinate (qout) at (1.10,2.00);

\coordinate (a4) at (2.10,0.55);
\coordinate (a3) at (2.10,-0.55);

\draw[fermion] (qin) -- (vtop) -- (qout);

\linearrow{qin}{vtop}{0.40}{0.63}
\linearrow{vtop}{qout}{0.37}{0.60}

\draw[gluon, line width=0.55pt] (gin) -- (vbot);
\draw[gluon, line width=0.55pt] (vbot) -- (vtop);

\filldraw (vbot) circle (0.065);

\draw[scalar] (vbot) -- (vh);
\node at (0.6,0.2) {$h$};

\filldraw (vh) circle (0.065);

\draw[scalar] (vh) -- (a4);
\draw[scalar] (vh) -- (a3);

\node at (-0.78,-0.7) {$g$};
\node at (-0.25,0.62) {$g$};

\node at (-0.8,2.00) {$q$};
\node at (0.8,2.00) {$q$};

\node at (1.80,0.60) {$a_i$};
\node at (1.80,-0.60) {$a_i$};

\end{scope}

\begin{scope}[xshift=4.40cm]

\coordinate (vbot) at (0,0);
\coordinate (vtop) at (0,1.25);
\coordinate (vh)   at (1.25,0);

\coordinate (gin)     at (-1.05,-0.60);
\coordinate (qbarin)  at (-1.10,2.00);
\coordinate (qbarout) at (1.10,2.00);

\coordinate (a4) at (2.10,0.55);
\coordinate (a3) at (2.10,-0.55);

\draw[fermion] (qbarin) -- (vtop) -- (qbarout);

\linearrow{vtop}{qbarin}{0.37}{0.60}
\linearrow{qbarout}{vtop}{0.40}{0.63}

\draw[gluon, line width=0.55pt] (gin) -- (vbot);
\draw[gluon, line width=0.55pt] (vbot) -- (vtop);

\filldraw (vbot) circle (0.065);

\draw[scalar] (vbot) -- (vh);
\node at (0.62,0.22) {$h$};

\filldraw (vh) circle (0.065);

\draw[scalar] (vh) -- (a4);
\draw[scalar] (vh) -- (a3);

\node at (-0.8,-0.7) {$g$};
\node at (-0.25,0.6) {$g$};

\node at (-0.8,2.00) {$\bar q$};
\node at (0.8,2.00) {$\bar q$};

\node at (1.80,0.60) {$a_i$};
\node at (1.80,-0.60) {$a_i$};

\end{scope}

\begin{scope}[xshift=8.80cm]

\coordinate (vbot) at (0,0);
\coordinate (vtop) at (0,1.25);
\coordinate (vh)   at (1.25,0);

\coordinate (g1) at (-1.05,-0.60);
\coordinate (g2) at (-1.10,2.00);
\coordinate (g5) at (1.10,2.00);

\coordinate (a4) at (2.10,0.55);
\coordinate (a3) at (2.10,-0.55);

\draw[gluon, line width=0.55pt] (g1) -- (vbot);
\draw[gluon, line width=0.55pt] (vbot) -- (vtop);
\draw[gluon, line width=0.55pt] (g2) -- (vtop);
\draw[gluon, line width=0.55pt] (vtop) -- (g5);

\filldraw (vbot) circle (0.065);

\draw[scalar] (vbot) -- (vh);
\node at (0.62,0.22) {$h$};

\filldraw (vh) circle (0.065);

\draw[scalar] (vh) -- (a4);
\draw[scalar] (vh) -- (a3);

\node at (-0.8,-0.70) {$g$};
\node at (-0.25,0.60) {$g$};
\node at (-0.8,2.10) {$g$};
\node at (0.8,2.10) {$g$};

\node at (1.80,0.60) {$a_i$};
\node at (1.80,-0.60) {$a_i$};

\end{scope}

\end{tikzpicture}
}

\caption{Leading Feynman diagrams for monojet production from the $O_{Ha}$ operator indicated  by a thick dot, with $q=u,d$ and $i=1,2$.}
\label{fig:monojet_diagrams_c_ha}
\end{figure}
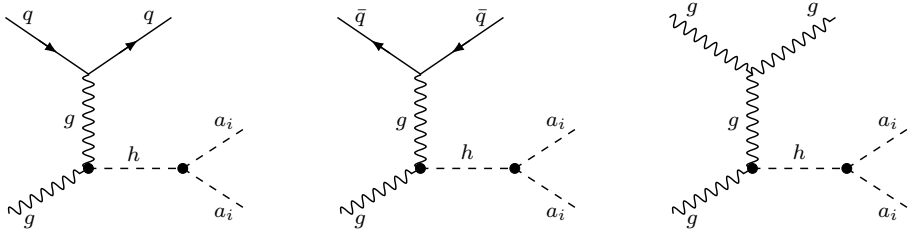

The LO operators in \cref{eq:DALP-LO-Lag} also contribute to monojet production at the LHC via the diagrams in  \cref{fig:monojet_diagrams_c_ba,fig:monojet_diagrams_c_ha}. We estimate the bounds on our effective operators from the ATLAS search reported in Ref.~\cite{ATLAS:2021kxv} (see also Ref.~\cite{CMS:2017zts} for the corresponding CMS analysis). The predicted number of events has been obtained implementing the operators under study in \textsc{FeynRules} and simulating monojet events with  \textsc{MadGraph5} \cite{Alwall:2014hca} with the PDF set PDF4LHC15\_nlo\_mc \cite{Butterworth:2015oua}, followed by parton showering and hadronization with \textsc{Pythia} \cite{Sjostrand:2006za,Sjostrand:2007gs}, and detector simulation using \textsc{Delphes3} \cite{deFavereau:2013fsa}. 

Our study follows a similar one performed in Ref.~\cite{Roy:2025pht} with some minor changes. In particular, we build the Poisson log-likelihood comparing our \textsc{MadGraph5} simulation for the expected signal events $s_i(c/f_a^2)$ for a coupling $c$  (i.e. $c_{Ba}$ or $c_{Ha}$), with the measured number $n_i$ and the expected SM background $b_i \pm \Delta_i$ reported in Table 8 of Ref.~\cite{ATLAS:2021kxv}. The uncertainty in the SM prediction is incorporated through nuisance parameters $\theta_i$ with the corresponding Gaussian pull $\Delta_i$.  One has 
\begin{equation}
    \chi^2(c/f_a^2,\theta_i)= \sum_i
2\left[\mu_i-n_i+n_i\ln\!\left(\frac{n_i}{\mu_i}\right)\right]+
\left(\frac{\theta_i}{\Delta_i}\right)^2\,,
\end{equation}
where $i$ labels the different bins in missing transverse energy, and where
\begin{equation}
    \mu_i=s_i(c/f_a^2)+b_i+\theta_i\,.
\end{equation}
The expected signal in each bin $i$ is
\begin{equation}
    s_i(c/f_a^2)=\mathcal{L}\,\sigma_{\rm fb}(c/f_a^2)\,\,(A \ \varepsilon)_i\,,
\end{equation}
where $\mathcal{L}$ is the integrated luminosity and $(A\ \varepsilon)_i$ is the acceptance times efficiency for that bin. After profiling over the nuisance parameters $\theta_i$, we derive the bounds reported in \cref{fig:EstudioH2,fig:EstudioZ2} imposing $\Delta
\chi_{\mathrm{tot}}^2(c/f_a^2)=2.71$,
which corresponds to a one-sided $95\%$ C.L. cut for one degree of freedom. 

When summing over the different bins in missing transverse energy, we ensure that $f_a/\sqrt{c} > E_T^{max}$, removing the highest energy bins as necessary to ensure the validity of the EFT. Our results correspond to the excluded purple areas in \cref{fig:EstudioH2,fig:EstudioZ2}.  In analogy to the monojet searches at ATLAS, the LO operators would lead to monophoton signatures at LEP~\cite{L3:2003yon,DELPHI:2003dlq}. However, these constraints would be weaker  given the lower energies probed. Finally, invisible dark photon searches, such as those performed by NA62~\cite{NA62:2019meo} with very competitive limits also would lead to rather weak constraints, given that they exploit the soft collinear enhancement of Bremsstrahlung emission, which would be absent for the operators under study.  

\subsection{Cooling bounds}

DALP emission inside supernovae can affect their cooling. Supernovae cooling bounds constrain a window of couplings: from the lower end, whenever DALPs are produced efficiently enough to affect stellar cooling, up to the trapping regime, whenever the coupling is so strong that DALPs cannot escape the supernova.  These constraints are only competitive for the $\mathcal{O}_{{Ba}}$ operator, since the universal  Higgs-mediated operator interactions are further suppressed by small Yukawa couplings~\cite{Bauer:2020nld} of the Higgs.

To estimate the supernovae cooling constraints on the $\mathcal{O}_{{Ba}}$ coefficient, we rescale the bounds derived in Ref.~\cite{DeRocco:2019jti} for the operator describing the interaction of an exotic fermion $\chi$ of mass $m_\chi$ with the electromagnetic current $J^{\mathrm{em}}_\mu$
\begin{equation}
\mathcal{O}_{\mathrm{vec}} = \frac{c_{\mathrm{vec}}}{\Lambda^2}\,\bar{\chi}\gamma^\mu \chi\, J^{\mathrm{em}}_\mu\,.
\label{eq:otherop}
\end{equation}
We follow the same procedure as Ref.~\cite{Fernandez-Martinez:2023phj}, i.e. to equate the production and capture cross sections induced by $\mathcal{O}_{\mathrm{vec}}$ with those induced by $\mathcal{O}_{Ba}$, at the boundaries of the $\mathcal{O}_{\mathrm{vec}}$ excluded region.\footnote{We have neglected the possible impact of the difference in the distribution of the bosonic DALPs versus the fermionic $\chi$ inside the SN. This effect should be small as, for relativistic particles, the corresponding energy densities only differ by a factor $7/8$, and the number density by $3/4$.}

In practice, since the cross sections depend on the supernova temperature as well as on the abundances of electrons, positrons and protons, which vary with the stellar radius, we consider the averaged cross sections, $\bar{\sigma}$,
\begin{equation}
    \bar{\sigma} = \frac{\int_{0}^{100\,\mathrm{km}} n(r)\, T(r)\, \sigma(T(r))\, dr}{\int_{0}^{100\,\mathrm{km}} n(r)T(r)\, dr} \, ,
    \label{4.13}
\end{equation}
where $T(r)$ and $n(r)$ are the temperature and particle density, respectively, both defined as functions of the distance from the center of the supernova $r$ reported in Ref.~\cite{DeRocco:2019jti}.

The lower bound is controlled by the DALP production rate which needs to be sufficient for the cooling  to be observable. The DALP production  is dominated by $e^-e^+$ annihilation and thus we impose:  
\begin{equation}
\textbf{lower bound:}\qquad \bar{\sigma}_{e^+ e^- \to \bar{\chi} \chi}(c_{\mathrm{vec}}/\Lambda^2,m_\chi)=
\bar{\sigma}_{e^+ e^- \to a_1 a_2}(c_{Ba}/f_a^2,m_a)\,, \quad \text{for\,\,} m_a=m_\chi\,,
\label{eq:lowerSN}
\end{equation}
where $c_{\mathrm{vec}}/\Lambda^2$ takes values along the lower bound frontier derived in Ref.~\cite{DeRocco:2019jti}.

The upper bound is instead determined by the strength of DALP scatterings with protons, electrons and positrons that may keep them trapped inside the supernova and prevent efficient cooling. Therefore, in  analogy with  \cref{eq:lowerSN}, to derive the upper bound on $c_{Ba}/f_a^2$ we require:
\begin{equation}
\textbf{upper bound:}\quad \left[\bar{\sigma}_{\chi e^{\pm} \to  \chi e^\pm}+\bar{\sigma}_{\chi p \to  \chi p}\right](c_{\mathrm{vec}}/\Lambda^2,m_\chi)= 
\left[\bar{\sigma}_{a_1 e^\pm \to  a_2 e^\pm}+\bar{\sigma}_{ a_1 p \to  a_2 p}\right](c_{Ba}/f_a^2,m_a)
\, ,
\end{equation}
for $m_a=m_\chi$ and 
where now $c_{\mathrm{vec}}/\Lambda^2$ takes values along the upper bound frontier  determined in Ref.~\cite{DeRocco:2019jti}.

The bounded area resulting from supernovae cooling for the $\mathcal{O}_{Ba}$ operator coefficient is shown in light blue in \cref{fig:EstudioZ2}.

\subsection{DALP relic abundance} \label{DALP_relic_abundance}

Given their conserved charge, DALPs are stable and therefore DM candidates. In the following, we study the associated allowed DALP DM parameter space. The freeze-out regime is excluded, except for a tiny resonance region in parameter space. The freeze-in regime is allowed for a large region of parameter space. Finally, the misalignment mechanism for DM production also works surprisingly well up to large DALP masses of order $10^{-1}\, \text{GeV}$ ($10^{-3}\, \text{GeV}$) for the hypercharge (Higgs) operator.  

Notice that, in principle, there is a fourth way to obtain the correct relic abundance for DM DALPs. Indeed, if self-interactions violating DALP number such as $3 \to 2$ processes are present, and potentially equilibrate within the DALP sector, their freeze-out could then control the final DALP abundance. For non-symmetric cosets, these odd-numbered DALP interactions are present in the DALP kinetic terms. Such self-interactions are invoked in the so-called strongly interacting massive particle (SIMP) DM models~\cite{Hochberg:2014dra,Hochberg:2014kqa}, although in that context they are introduced via a Wess-Zumino term, which is smaller than the kinetic terms by two derivatives and a loop factor of $1/(24\pi^2)$. In principle, self-interactions would alter the predictions of freeze-out and freeze-in from the SM bath (see also \cite{Davighi:2024zip,Davighi:2025awm}). Nevertheless, we do not include them in our analyses, as here we focus in constraining the LO operators mediating SM-DALP interactions and the potential presence of the odd-numbered DALP interactions introduces additional model dependence.
We examine the freeze-out, freeze-in and misalignment scenarios in turn.

\subsubsection*{Freeze-out}

Since the interaction rate of a $d=6$ operator grows faster with temperature than the Hubble rate, there is always a decoupling temperature  $T_\text{dec}\sim \big({f_a^4}/{(C^2 M_\text{Pl})}\big)^{1/3}$~\cite{Weinberg:2013kea}, where $C= c_{Ba} \  \text{or} \ c_{Ha}$, above which DALPs are in thermal equilibrium with the SM bath. Whenever DALPs are  light enough to be produced at reheating and the reheating temperature is larger than the decoupling temperature, $m_a< T_\text{dec} <  T_R$, DALPs are produced via freeze-out.  In~\cref{fig:EstudioH2,fig:EstudioZ2},  the solid black line indicates     the values of the DALP mass and its coupling for each of the operators for which the correct relic abundance would be obtained. It illustrates the near impossibility for DALPs to reproduce the correct dark matter abundance via freeze-out. The computation has been done with the \textsc{MicrOMEGAs}~\cite{Belanger:2001fz, Alguero:2023zol} code. The ``throats'' in both plots correspond to the Higgs boson and $Z$ resonances, respectively, since DALP production has contributions with those particles in the $s$-channel. Moreover, for the plot in \cref{fig:EstudioH2} involving the operator with the Higgs boson, the steps with abrupt changes correspond to kinematic thresholds for SM particles producing DALPs in the early universe, e.g. $W^+ W^-, ZZ...$, etc. Since the Higgs couplings are proportional to the particle mass, these thresholds turn out to be very significant contributions, while their impact is much less noticeable in \cref{fig:EstudioZ2} for the operator coupled to hypercharge. For $m_a \lesssim 50$~GeV, the line  stops as no value of $c_{Ha}$ can account for the correct relic abundance via freeze-out, even disregarding the experimental constraints. Indeed, as $m_a$ decreases, the DALP annihilation cross section needs to increase to compensate the lower $T_\text{dec}$ so as to keep the DALP relic energy density constant, and thus $c_{Ha}/f_a^2$ must grow.  However, the cross section growth with $c_{Ha}/f_a^2$ moderates when approaching $ c_{Ha}/f_a^2  \sim 1/v^2$ since the operator also introduces a correction to the DALP kinetic terms that propagates to the physical coupling upon field renormalization ($Z_a$ in \cref{hdecay}). Nevertheless, that region is already beyond the validity of the EFT.

\subsubsection*{Freeze-in}

For couplings smaller than the one required to obtain the correct relic abundance via freeze-out, the DALPs will decouple too early\footnote{Often the decoupling occurs when the DALPs are still relativistic, since the production rate is enhanced with respect to the Hubble expansion at higher $T$, being mediated by effective $d=6$ operators.} and overclose the Universe. For even smaller couplings, $T_R<T_\text{dec}$, DALPs never reached  thermal equilibrium, which is the freeze-in production regime. The transition between the freeze-in and freeze-out regimes and the prediction for their final relic abundance is significantly model-dependent. Indeed, for interactions mediated by $d=6$ operators, the relic abundance grows with the third power of the reheating temperature $T_R$~\cite{Blennow:2013jba}.  In~\cref{fig:EstudioH2,fig:EstudioZ2}  for the $\mathcal{O}_{Ha}$ and $\mathcal{O}_{Ba}$ operators, respectively,  the different orange lines --labeled by the value of $T_R$--  correspond to the values of the parameters for which the correct relic abundance would be obtained via freeze-in. For a given value of $T_R \gg m_a$, the area between freeze-out and freeze-in would overproduce the DM abundance and is excluded. 
  
Nevertheless, if $T_R \lesssim m_a$, DALP production would start when already Boltzmann suppressed~\cite{Goudelis:2026lyy,Bernal:2026clv}. In this regime, the correct relic abundance could be obtained for sufficiently large couplings compensating this suppression. This behaviour can also be seen  in~\cref{fig:EstudioH2,fig:EstudioZ2}. For a given value of $T_R$, when $m_a \gtrsim T_R$ the coupling required grows exponentially with $m_a$ to compensate the Boltzmann suppression. Conversely, for $T_R \gtrsim m_a$, the required coupling decreases with increasing $m_a$ and, for the operator coupling DALPs to hypercharge, roughly scales with $T_R^{-1.5}$, as expected from the $T_R^3$ dependence of the relic abundance:  this is shown in~\cref{fig:EstudioZ2} by the approximately equidistant lines for different $T_R$ values.  This scaling is instead violated for the coupling to the Higgs since, when lowering $T_R$, the more massive SM bath particles will subsequently also become Boltzmann suppressed and significantly reduce the cross section, as the Higgs couples to mass. Hence, even larger couplings are required to compensate, as shown in~\cref{fig:EstudioH2} by the much larger distance for the $T_R=1$~GeV line. On the other hand, we interpret the shorter distance  between the $T_R=10^3$~GeV and $T_R=10^2$~GeV lines in this figure as due to the fact that the latter is close to Higgs resonance and hence smaller couplings are required to compensate.  Indeed, in the regime $m_a<T_R$ and for reheating temperatures above the electroweak scale, they can be well reproduced by an analytic estimation~\cite{Blennow:2013jba}:
\begin{equation}
   \Omega_{\rm DALP} h^2 \simeq
\kappa_{X}
\left[
3 \times 10^{24}\,
m_a\,
\frac{1}{9}\,
\frac{\mathcal{B}^6}{g_{*s}(T_R)\sqrt{g_*(T_R)}}\,
T_R^3
\left(\frac{c_X}{f_a^2}\right)^2
\right]
\label{eq:freeze_in_analytic}
\end{equation}
with  $\mathcal{B}=3.55$ and where $X=Ha, Ba$. The coefficient $\kappa_{X}$ depends on the dominant production cross section for each operator. In particular, $\kappa_{Ha}={1}/({32\pi)}$ and  $\kappa_{Ba}={5g_1^2}/{(12\pi)}$ for the Higgs ($\mathcal{O}_{Ha}$) and hypercharge ($\mathcal{O}_{Ba}$) operators, respectively. With these values, the analytic expression reproduces the numerical freeze-in abundance line with a relative error around $20 \%$ for the Higgs operator and  below $1 \%$ for the hypercharge operator. The discrepancy in the Higgs case arises because in Ref.~\cite{Blennow:2013jba} the parameter $\mathcal{B}$ was adjusted for production from SM fermions in the thermal bath, whereas in our case the dominant production is via the Higgs boson. The difference in the thermal statistics can account for the discrepancy. Fitting the coefficient to our numerical freeze-in line instead gives  $\mathcal{B} =3.72$.

The freeze-in abundance generated by $\mathcal{O}_{Ba}$ was already considered in Ref.~\cite{Dessert:2025yvk} in the context of multiple axions from $U(1)^N$ breakings, although $d=5$ operators were also present, leading to a different phenomenology.

\subsubsection*{Misalignment}

DALPs, which are weakly coupled massive scalars, can also be produced via the misalignment mechanism~\cite{Preskill:1982cy,Abbott:1982af,Dine:1982ah}.  For a standard ALP with a $d=5$ coupling to the electromagnetic field strength such as $aF\widetilde{F}/f_a$, ALP decay into two photons is severely constrained~\cite{Langhoff:2022bij,Arias:2012az}, which excludes the ALP as a DM candidate for $m_a\gtrsim\mathcal{O}(10^2\text{--}10^3)\,\text{eV}$. DALPs are different, though. They carry a conserved charge under the unbroken dark symmetry group $H$, so decays of the lightest DALP are forbidden. In our degenerate DALP case, all of them are stable. The usual ALP constraints simply do not apply, and misalignment can in principle populate the DALP dark matter band at much larger masses, as we proceed to discuss.

The misalignment mechanism for DALP dark matter proceeds  alike to that for ordinary ALPs. At high temperatures, the DALP field is frozen at some initial angle $\theta_i$ by Hubble friction until the oscillations begin, when $H(T_\text{osc})\sim m_a$ at $T_\text{osc}\sim\sqrt{m_a M_\text{Pl}}$. From then on the DALP oscillates around its minimum and these coherent oscillations behave as cold dark matter, with its energy density redshifting as $(1+z)^{3}$. The resulting relic density for \emph{each} DALP field reads\footnote{The numerical coefficient differs slightly from some results in the literature due to a different choice for the onset of oscillations. We use $1.67\,H = m_a$ rather than $3H = m_a$, which gives a better fit to the numerical solution~\cite{AlonsoAlvarez:2019cgw,DiLuzio:2021gos}.}
\begin{equation}
\frac{\rho_{a,0}}{\rho_\text{DM}} \simeq 5.7 \sqrt{\frac{m_a}{1\, \text{eV}}} \left(\frac{\theta_i f_a}{10^{12}\,\text{GeV}}\right)^2 \mathcal{F}(T_\text{osc}) \,,
\label{eq:misalignment_relic}
\end{equation}
where $\mathcal{F}(T_{\rm osc}) \equiv\left(g_{*}(T_{\rm osc}) / 3.38\right)^{\frac{3}{4}}\left(g_{s}(T_{\rm osc}) / 3.93\right)^{-1}$ is an $\mathcal{O}(1)$ factor and $\rho_{\rm DM}\simeq 1.26\, {\mathrm{keV}}/{\mathrm{cm}^{3}}$ from Planck 2018 data~\cite{Aghanim:2018eyx}.  In using \cref{eq:misalignment_relic}, we are assuming that the Universe reached $T_{\rm osc}$ at some point, so $T_R>T_{\rm osc}\sim \sqrt{m_a M_{\rm Pl}}$. 

\begin{figure}
    \centering
    \centering
    \includegraphics[width=\textwidth]{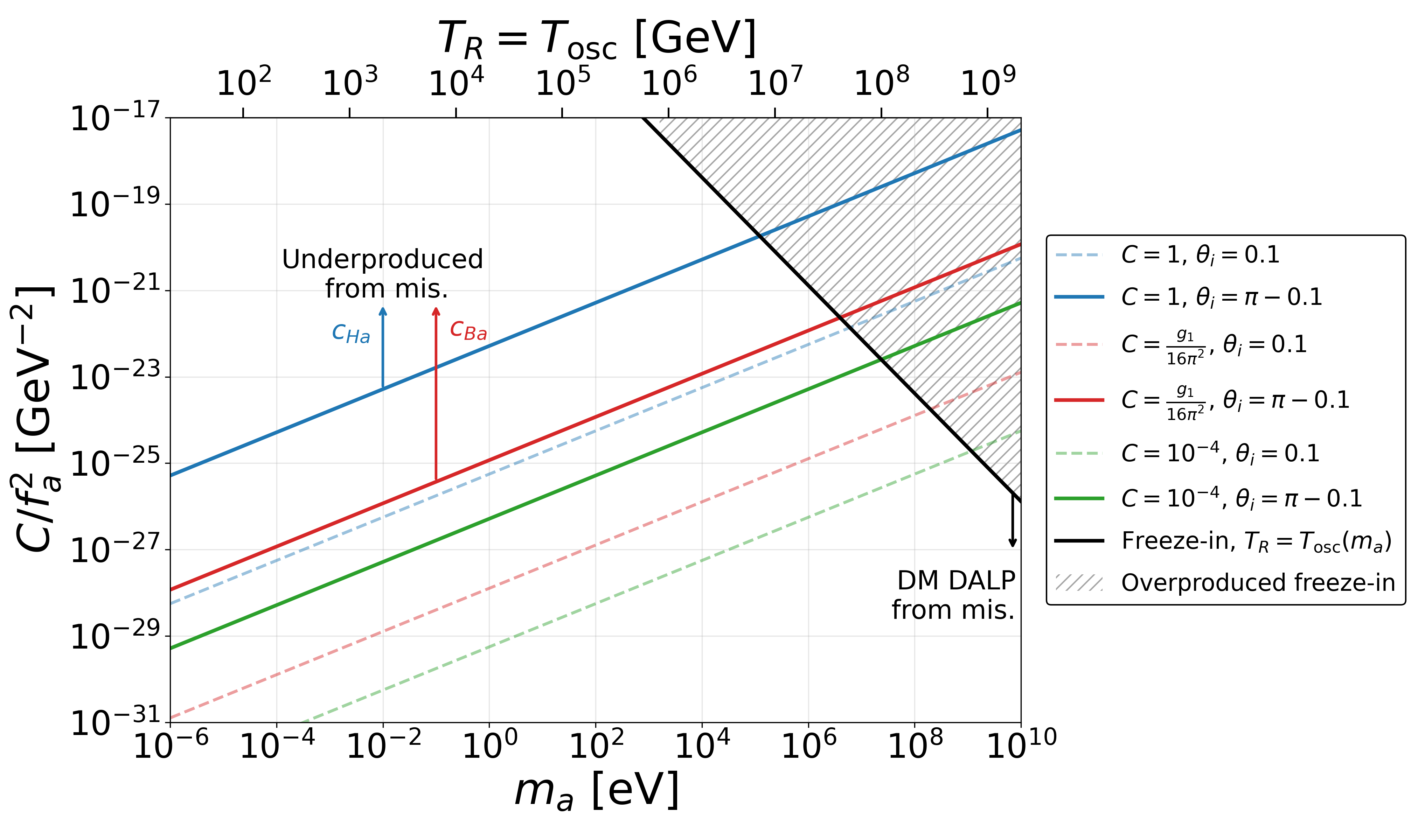}
     \caption{\small  Parameter space for DALP dark matter produced via misalignment in the $(m_a,C/f_a^2)$ plane with $C=\{c_{Ha},c_{Ba}\} $. The coloured lines correspond to representative choices of $C$ and $\theta_i$ that reproduce the DM abundance via misalignment. The black hatched region corresponds to freeze-in overproduction for $T_R=T_{\rm osc} (m_a)$, computed for the hypercharge operator. Using  the  Higgs operator instead would modify the line by about $14\%$, as discussed below \cref{eq:freeze_in_analytic}. 
     In the region above the solid blue and red lines, the misalignment abundance is subdominant for couplings satisfying the NDA bound $c_{Ha}\lesssim 1$,  $c_{Ba}\lesssim {g_1}/(16 \pi^2)$, respectively. Conversely, below those lines there is always a choice of the initial misalignment angle and a value of $C$ within its NDA range that leads to the correct DM abundance. 
     \label{fig:misalignment}
 }
\end{figure}

Setting the ratio in Eq.~(\ref{eq:misalignment_relic}) to one for fixed $\theta_i$ gives a curve in the $(1/f_a^2,m_a)$ plane where DALPs explain 100\% of the DM relic density. Curves for various $\theta_i$ values are plotted in \cref{fig:misalignment}, for $c_{Ha}/f_a^2$ and $c_{Ba}/f_a^2$ as a function of $m_a$. Note that  the maximal value depicted for the coefficients differs: $c_{Ha}=1$ versus $c_{Ba}=g_1/(16\pi^2)$, as mandated by naive dimensional analysis (NDA) \cite{Manohar:1983md,Gavela:2016bzc}. Indeed, while we have used in this paper the conventional definition of the operator coefficients, NDA would suggest instead the parametrization
\begin{equation}
\mathcal{L}_{\rm{LO}}^{d=6}= \frac{\hat{c}_{Ha}}{f_a^2}\,   \mathcal{O}_{Ha}  \,+\,  \frac{g_1 \,\hat{c}_{Ba}}{16\pi^2\,f_a^2}\,  \mathcal{O}_{Ba},
\label{eq:DALP-LO-Lag-NDA}
\end{equation}
where  $g_1$ denotes the SM $U(1)_Y$ hypercharge gauge coupling constant, and $\hat{c}_{Ha}\lesssim 1$, $\hat{c}_{Ba}\lesssim 1$ to remain in the perturbative regime. Comparing with Eq.~(\ref{eq:DALP-LO-Lag}), it follows that ${c}_{Ha}=\hat{c}_{Ha}$ while  ${c}_{Ba}=g_1 \hat{c}_{Ba}/(16\pi^2)$.

How heavy can a misalignment DALP DM candidate be?  One may assume that the key question is whether the zero mode (the homogeneous component of the field) survives or gets thermalized by the SM bath. Even without $d=5$ operators, the $d=6$ interaction could in principle thermalize the DALP population. 
However, one can check that for $d=6$ operators the main effect that prevents having heavier DALPs from misalignment is overproduction via freeze-in. As we discussed in \cref{eq:freeze_in_analytic}, values of the coupling above the black freeze-in line in \cref{fig:misalignment} result in a DALP relic density larger than the DM one and thus they are excluded. Nevertheless, it is interesting to note that DALPs can be DM up until surprisingly large masses for a misalignment relic, as compared to the customary singlet ALPs of the $d=5$ Lagrangian.  For example, for $\theta_i=0.1$, the DALP misalignment region extends until $m_a\sim 200$ MeV for $c_{Ba}\sim g_1/(16\pi^2)$ and until  $m_a\sim 6$ MeV for $c_{Ha}\sim 1$.

\subsection{Direct searches} \label{Direct_searches}

If DALPs constitute the observed DM of the universe, they would lead to signatures in direct DM search experiments such as LZ~\cite{LZ:2024zvo} from DALP-nucleon scattering.

For the first operator in \cref{eq:DALP-LO-Lag}, $\mathcal{O}_{Ha}$, its interaction with nucleons will be through the effective Higgs-nucleon coupling (see e.g.~\cite{DelNobile:2021wmp})
\begin{equation}
\mathcal{L}_{hNN} \;=\; -\, g_{hNN}\, h\,\bar N N,
\qquad
g_{hNN} \;=\; \frac{m_N}{v}\, f_N,
\label{eq:ghNN}
\end{equation}
where $m_N$ is the nucleon mass and $f_N= 0.30 \pm 0.03$ parametrizes the nucleon matrix element~\cite{Hoferichter:2015dsa}. For the squared amplitude for DALP-nucleon scattering, we find
\begin{align}
|\mathcal M(t)|^2
&=\left(\frac{2f_N\,m_Nc_{Ha}}{f_a^2Z_a}\right)^{2}
\frac{\bigl(m_a^2-\tfrac{1}{2}t\bigr)^2\,(4m_N^2-t)}{(t-m_h^2)^2}.
\end{align}
In the nonrelativistic limit, for each DALP species, we approximate $t\to 0$ and $s \simeq (m_a+m_N)^2$, obtaining
\begin{align}
\sigma_{aN}^{\rm SI}(0)
&\simeq \frac{|\mathcal M(0)|^2}{16\pi\,s}
\simeq  \frac{f_N^2\,m_N^4\,m_a^4c_{Ha}^2}{\pi\,f_a^4\,m_h^4\,(m_a+m_N)^2Z_a^2}.
\end{align}

For the second operator in \cref{eq:DALP-LO-Lag}, $\mathcal{O}_{Ba}$, the scattering cross section is dominated by photon exchange. The photon–nucleon coupling reads\cite{DelNobile:2021wmp}:
\begin{equation}
\langle N'|J^\mu_{\rm EM}|N\rangle_{\rm NR}
= \begin{cases}
2m_N Q_N\,I_N, & \mu=0 \\
Q_N\,\left( \mathbf k+\mathbf k' \right)^i \,I_N
+i\,g_N\, [ \mathbf s_N\times\left( \mathbf k-\mathbf k' \right)]^i, & \mu = i
\end{cases}
\end{equation}
where $\mathbf k$ and $\mathbf k'$ are the nucleon tri-momenta, $Q_N$ is the nucleon charge ($Q_p=1$, $Q_n=0$), $g_N$ its effective magnetic moment, and $I_N$ is the identity matrix in nucleon spin space. Keeping only the spin–independent part and neglecting the velocity-suppressed $Q_N$ term, we obtain at LO
\begin{equation}
\mathcal M_{N,{\rm SI}}^{\rm NR}
=
2e\, c_W\frac{c_{Ba}}{f_a^2}\, m_N\,Q_N \frac{s+m_a^2-m_N^2}{\sqrt{s}}. \label{ec.Direct_searches_photon}
\end{equation}
Using $s \simeq (m_a+m_N)^2$, it follows that
\begin{equation}
\sigma_{\rm SI}
=
\frac{e^2\,c_{Ba}^{2}\,c_W^{2}\,m_N^2\,m_a^2}{\pi\,f_a^{4}(m_a+m_N)^2}\;
\;
Q_N^2.
\end{equation}

The comparison of the cross sections obtained above with the constraints obtained in Ref.~\cite{LZ:2024zvo} allows one to derive the bounds from direct searches, which are depicted in \cref{fig:EstudioH2,fig:EstudioZ2} as excluded green regions. Future prospects from XLZD with an exposure of 1000~\text{ton-yr}~\cite{XLZD:2024nsu} are depicted by the hatched green area. Notice that the values of the couplings that lead to the correct relic abundance via freeze-out production are already excluded by the combination of constraints from direct searches at LZ, invisible widths, and monojet constraints at ATLAS.\footnote{Except for a very small area close to the Higgs resonance for $2m_a=m_H$.} On the other hand, the freeze-in region in which DALPs constitute the observed DM is challenging to probe given the very feeble couplings required unless $T_R \lesssim m_a$.

\section{Examples of UV completions and generalization}\label{sec:uvcompletions}

We present next some perturbative ultraviolet (UV)  complete models containing heavy scalars and/or fermions, in order to illustrate how the lowest order DALP Lagrangian \cref{eq:DALP-LO-Lag} can be generated, with particular emphasis on the $d=6$ antisymmetric DALP operator $\mathcal{O}_{Ba}$. 

The simplest example is an unbroken $SO(2)$ symmetry, i.e. a dark $U(1)$, which can arise from a spontaneously broken $SO(3)$ symmetry in the dark sector, through the breaking 
\begin{equation}
SO(3) \longrightarrow SO(2)\, .
\end{equation}
The GBs of this symmetry breakdown live on the coset manifold 
$SO(3)/SO(2)\simeq S^2$.
Let
$$
 \Phi^i=(\Phi^1,\Phi^2,\Phi^3)
 $$
be a scalar field, which is a singlet of the SM gauge symmetry and a triplet of $SO(3)$.
The potential $V(\Phi)$ can be written as
\begin{equation}
V(\Phi) = \frac{\lambda_\Phi}{4}\left(\Phi^i\Phi^i - f_a^2\right)^2\,.
\label{Phi-pot}
\end{equation}
The potential is minimized by a vacuum expectation value of $\Phi$ which breaks the $SO(3)$ symmetry to $SO(2)$. The radial scalar singlet mode $\rho$ has mass $m_\rho^2 = 2 \lambda_\Phi f_a^2$, and the two angular GB modes are the DALPs $a_i$, $i=1,2$.

We explore below four sample illustrative UV completions.  The cases are:
\begin{itemize}
\item  The dark sector consists of only the $\Phi$ field, and its interaction with the SM is through the portal coupling to the SM Higgs doublet, 
\begin{equation}
\mathcal{L}= -\lambda_{\Phi H}\, (H^\dagger H) (\Phi^T \Phi)\,.
\end{equation}
\item The dark sector consists of $\Phi_i$ and additional heavy dark fermions $\Psi_i$, which are {\emph{triplets}} of $SO(3)$ and carry SM hypercharge  $Y_F$.  The dark scalars and fermions interact through the dark Yukawa interaction
\begin{equation}
\mathcal{L}=iy \,\epsilon_{ijk}\,\bar \Psi_i \Psi_j\Phi_k\,.
\label{singlet-fermionUV}
\end{equation}
\item The dark sector consists of $\Phi_i$ and additional heavy dark fermions $\Psi_i$, which are {\emph{doublets}} of $SO(3)$ and carry SM hypercharge $Y_F$.  The dark scalars and fermions interact through the dark Yukawa interaction
\begin{equation}
\mathcal{L}=y \,\bar \Psi  \tau^i  \Psi \,  \Phi_i \,.
\label{triplet-fermionUV}
\end{equation}
\item The dark sector consists of $\Phi_i$ and additional heavy dark scalars $S_i$, which are triplets of $SO(3)$ and carry hypercharge $Y_S$.  The dark scalars interact through the dark interaction
\begin{equation}
\mathcal{L}=i\mu \,\epsilon_{ijk}\, S_i^{*}   S_j\Phi_k\,,
\label{scalarUV}
\end{equation}
where $\mu$ has dimensions of mass.
\end{itemize}
Generalizations of the above examples to other $G \to H$ symmetry breakings are given in \cref{sec:uvcompletions2}.

\subsection{The universal  DALP operator \boldmath{$\mathcal{O}_{Ha}$}}
\label{subsec:UniversalDALPOp}
In the presence of the real scalar $\Phi$, the most general renormalizable Lagrangian will contain a portal interaction with the Higgs doublet,
\begin{align}
\mathcal{\delta L} &
 \supset \frac12 (\partial_\mu \Phi_i) (\partial^\mu \Phi_i) -  V(\Phi) -  \lambda_{\Phi H}  (H^\dagger H) (\Phi^T \Phi)  \,.
\label{eq:Universal-loop}
\end{align}
The DALPs are generated when $\Phi$ develops a vacuum expectation value $\vev{\Phi}=(0,0,f_a)^T$.

In the broken phase,
\begin{align}
\Phi &= (f_a + \rho)\, \xi \begin{bmatrix} 0 \\ 0 \\ 1\end{bmatrix} \, , & \xi &= e^{i \bm{a} /f_a} \, , 
& \bm{a} &= -i\begin{bmatrix} 0 & 0 & a_1 \\ 0 & 0 & a_2 \\
-a_1 & -a_2 & 0 \end{bmatrix} \, ,
\label{3.8}
\end{align}
where $\rho$ is the radial mode, and DALPs are angular modes, with $\xi$ being the chiral field. The scalar $\rho$ is massive, with $m_\rho^2=2 \lambda_\Phi f_a^2$, whereas the DALPs are massless GBs.

The contribution to SMEFT operators resulting from integrating out the radial scalar $\rho$ is well known;   tree-level $\rho$ exchange (see \cref{fig:tree}) yields the effective operator
\begin{figure}
\begin{center}
\begin{tikzpicture}
\draw[dashed] (0,0) -- (2,0);
\draw (1,0.25) node {$\rho$};
\draw[dashed,-Latex] (2,0) -- +(45:0.9);
\draw[dashed] (2,0)+(45:0.9) -- +(45:1.5);

\draw[dashed,-Latex] (0,0) -- +(135:0.9);
\draw[dashed] (0,0)+(135:0.9) -- +(135:1.5);

\draw[dashed] (2,0) -- +(-45:0.8);
\draw[dashed,-Latex] (2,0)+(-45:1.5) -- +(-45:0.8);

\draw[dashed] (0,0) -- +(225:0.8);
\draw[dashed,-Latex] (0,0)+(225:1.5) -- +(225:0.8);

\draw (2,0)+(45:2) node {$H$};
\draw (2,0)+(-45:2) node {$H$};
\draw (0,0)+(225:2) node {$H$};
\draw (0,0)+(135:2) node {$H$};
\draw(-0.75,0) node {$\lambda_{\Phi H}$};
\draw(2.75,0) node {$\lambda_{\Phi H}$};

\filldraw (2,0) circle (0.075);
\filldraw (0,0) circle (0.075);

\draw (1,-2) node {(a)};

\end{tikzpicture}
\hspace{2cm}
\begin{tikzpicture}
\draw[dashed] (0,0) -- (2,0);
\draw (1,0.25) node {$\rho$};
\draw[dashed] (2,0) -- +(45:1.5);

\draw[dashed,-Latex] (0,0) -- +(135:0.9);
\draw[dashed] (0,0)+(135:0.9) -- +(135:1.5);

\draw[dashed] (2,0) -- +(-45:1.5);

\draw[dashed] (0,0) -- +(225:0.8);
\draw[dashed,-Latex] (0,0)+(225:1.5) -- +(225:0.8);

\draw (2,0)+(45:2) node {$a_i$};
\draw (2,0)+(-45:2) node {$a_i$};
\draw (0,0)+(225:2) node {$H$};
\draw (0,0)+(135:2) node {$H$};
\draw(-0.75,0) node {$\lambda_{\Phi H}$};

\filldraw (1.925,-0.075) rectangle (2.075,0.075);
\filldraw (0,0) circle (0.075);

\draw (1,-2) node {(b)};

\end{tikzpicture}
\end{center}
\caption{ \label{fig:tree}Graphs contributing to (a) the $\mathcal{O}_{H\Box}$ operator and (b) The $\mathcal{O}_{Ha}$ operator. The solid circle is the $\lambda_{\Phi H} $ interaction, and the solid square is the interaction from the $\Phi$ kinetic term.}
\end{figure}
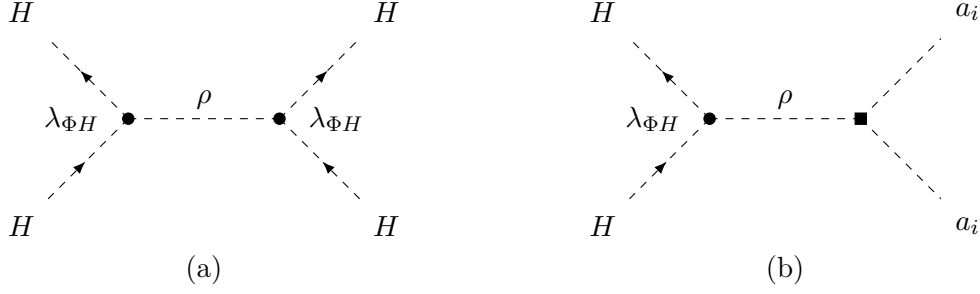
\begin{align}
    \mathcal{L}_{\text{eff}} \supset  \frac{C_{H\Box}}{f_a^2} \,\mathcal{O}_{H \Box} \,,
    \label{Leff}
\end{align}
where
\begin{align}
\mathcal{O}_{H \Box} \equiv (H^\dagger H) \Box (H^\dagger H)\,, \qquad \qquad C_{H\square} &= -\frac{\lambda_{\Phi H}^2}{2 \lambda_\Phi^2}\,.
\label{5.10}
\end{align}
This operator impacts the Higgs interactions, leaving its signature in the physical cubic and quartic Higgs self-coupling.  
The observational constraints and future projected sensitivities for $C_{H\Box}$ are summarized in Table~\ref{tab:CHbox-bounds}. 
\begin{table}
    \centering
    \begin{tabular}{|l|c|c|}
        \hline
        Present observations &   $-1.879 < C_{H\Box}\left(1\,\mathrm{TeV}/f_a\right)^2 < 1.715$ \quad ($95\%$) \\
        \hline
        HL-LHC 
         & $ |C_{H\Box}|\left(1\,\mathrm{TeV}/f_a\right)^2 < 0.72$ \quad ($95\%$)\\
        \hline
        FCC-ee
         & $ |C_{H\Box}|\left(1\,\mathrm{TeV}/f_a\right)^2 < 0.070$ \quad ($95\%$) \\
        \hline
    \end{tabular}
\caption{Present and projected constraints on the dimensionless combination $C_{H\Box}(1\,\mathrm{TeV}/f_a)^2$. The present bound is the linear-marginalised $95\%$ SMEFiT interval, while the HL-LHC and FCC-ee projection is obtained from the corresponding global fit of the linear-marginalised bound in Ref.~\cite{Celada:2024mcf}.  
    \label{tab:CHbox-bounds}}
\end{table}
\footnotetext{Note that Ref.~\cite{Enguita:2025ybx} incorrectly cites a precision of $0.028$. A recheck of the results shows that they were obtained assuming a precision $0.034$, in agreement with Ref.~\cite{Mangano:2020sao}.}

\begin{figure}
\begin{center}
\includegraphics[width=14cm]{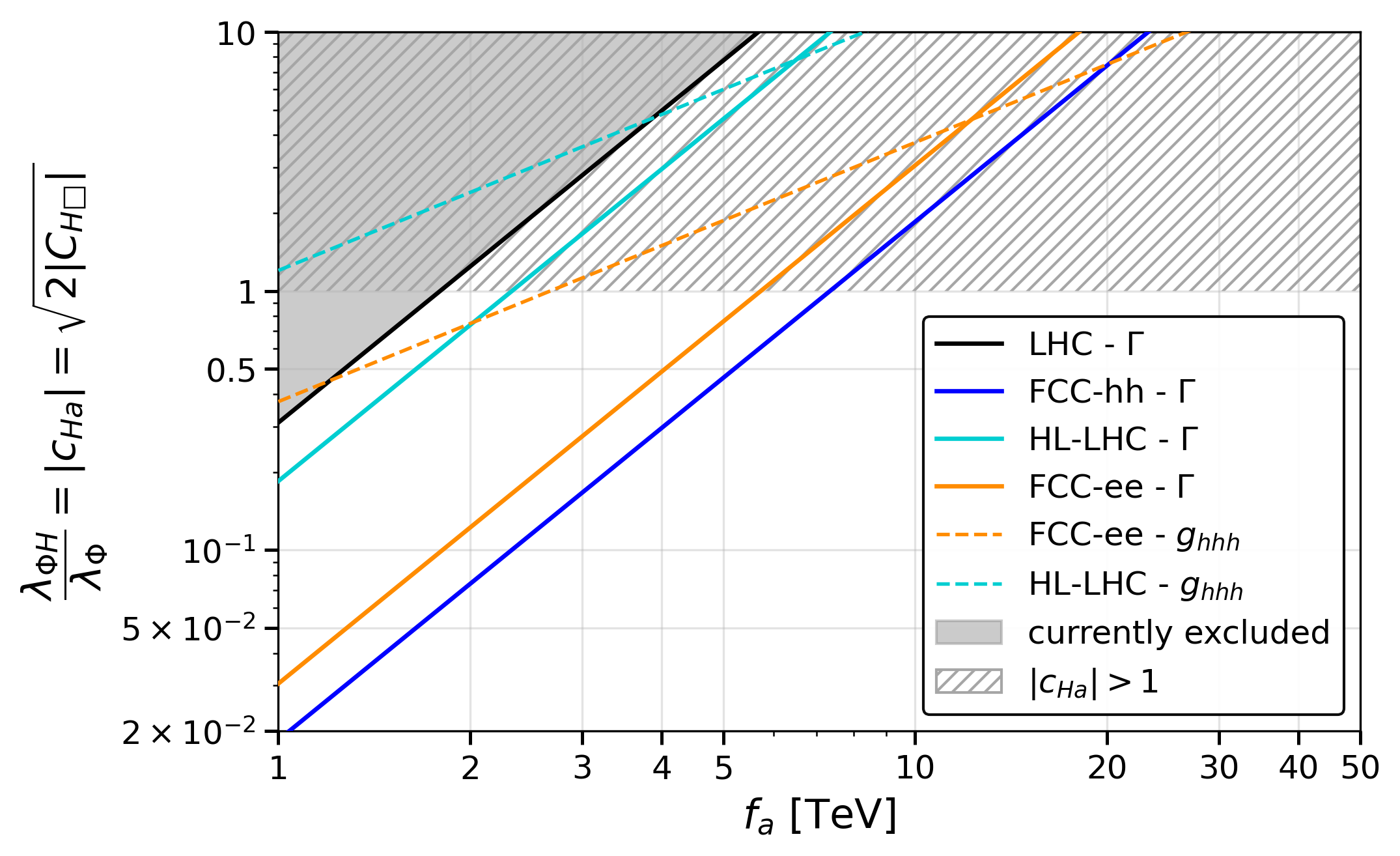}
\end{center}
\caption{The $\{\lambda_{\Phi H}/\lambda_\Phi, f_a\}$ plane showing relevant collider reach from LHC (black), FCC-hh (blue) and FCC-ee (orange). Solid lines represent bounds from the  invisible Higgs width in ~\cref{tab:CH-inv}, evaluated in the $m_a \to 0$ limit, while dashed lines the bounds on $C_{H\Box}$ based on the triple-Higgs coupling coefficient $g_{hhh}$ listed in Table~\ref{tab:CHbox-bounds}. The shaded region is excluded by current LHC data. The NDA inequality is $\left| c_{Ha} \right| \lesssim 1$.}
\label{fig:cHacHBox}
\end{figure}

Tree-level exchange of the heavy $\rho$ also contributes to the universal DALP interaction $(H^\dagger H) \partial_\mu a_i \partial_\mu a_i$, i.e.\ the first operator in \cref{eq:DALP-LO-Lag}, through the second graph in \cref{fig:tree},
\begin{align}
\mathcal{L} &= - \frac{\lambda_{\Phi H}}{\lambda_\Phi} (H^\dagger H)  u_i^\mu u_i^\mu 
= - \frac{\lambda_{\Phi H} }{\lambda_\Phi f_a^2} (H^\dagger H)  (\partial_\mu a_i )(\partial_\mu a_i)\,,
\label{tree-level-universal}
\end{align}
so that the $\mathcal{O}_{Ha}$ coupling is
\begin{align}
\frac{c_{Ha}}{f_a^2} &= - \frac{\lambda_{\Phi H} }{\lambda_\Phi f_a^2} \,.
\label{tree-level-universal2}
\end{align}
$C_{H\Box}$ in \cref{5.10} and $c_{Ha}$ in \cref{tree-level-universal2} depend on ${\lambda_{\Phi H} }/{\lambda_\Phi}$ and, in this model,
\begin{align}
\lvert c_{Ha} \rvert&= \sqrt{2 \lvert C_{H\Box} \rvert} \,.
\end{align}
The invisible width constraint on $c_{Ha}$ is given in \cref{sec:invwidth}. The triple Higgs coupling $g_{hhh}$ constrains $C_{H\Box}$.
The reach of the HL-LHC and FCC-ee in the $\{{\lambda_{\Phi H} }/{\lambda_\Phi},  f_a$\} plane is shown in \cref{fig:cHacHBox}. FCC-hh is less sensitive than FCC-ee to this parameter~\cite{Mangano:2020sao}.

These  facilities can detect deviations from the SM if the couplings are in the region above the corresponding lines.  While the values shown in \cref{fig:cHacHBox} are for the model of \cref{eq:Universal-loop}, the general form of the constraint will extend to other models.

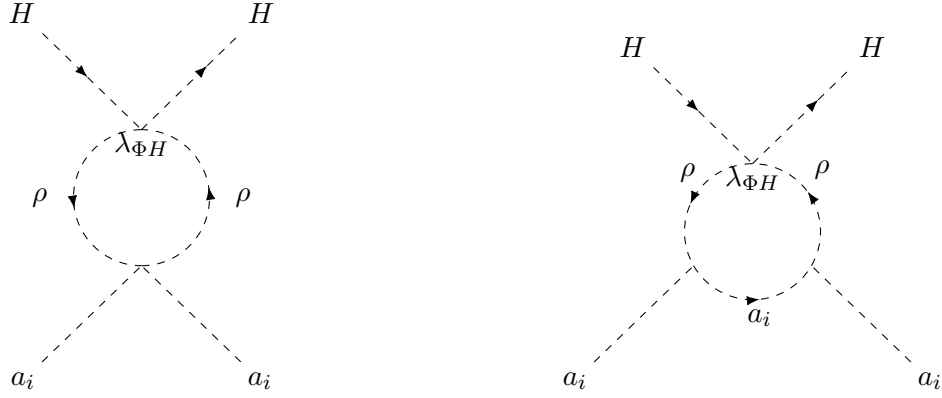
\begin{figure}[htbp]
\centering

\begin{subfigure}{0.48\textwidth}
\centering
\begin{tikzpicture}[scale=0.75]
  \begin{feynman}

    \def\r{1.2}
    \coordinate (O) at (0,0);

    \coordinate (vH) at ($(O)+(90:\r)$);
    \coordinate (v1) at ($(O)+(270:\r)$);
    \node at ($(vH)-(0,0.25)$) {\(\lambda_{\Phi H}\)};

    \vertex [above left=1.8cm of vH]  (Hin)  {\(H\)};
    \vertex [above right=1.8cm of vH] (Hcin) {\(H\)};

    \vertex [below left=1.8cm of v1]  (a1) {\(a_i\)};
    \vertex [below right=1.8cm of v1] (a2) {\(a_i\)};

    \diagram*{

      (a1)   -- [scalar] (v1),
      (a2)   -- [scalar] (v1),
    };

    \draw[dashed,-Latex](Hin) -- ($(Hin)!0.55!(vH)$);
    \draw[dashed] ($(Hin)!0.55!(vH)$) -- (vH);

    \draw[dashed,-Latex] (vH) -- ($(vH)!0.55!(Hcin)$);
    \draw[dashed] ($(vH)!0.55!(Hcin)$) -- (Hcin);
        
    \tikzset{
      looparrow/.style={
        postaction={decorate},
        decoration={markings, mark=at position 0.55 with {\arrow{Latex}}}
      },
      Sprop/.style={dashed}
    }

    \draw[Sprop, looparrow]
      (O) ++(90:\r) arc (90:270:\r)
      node[pos=0.5, left, xshift=-6pt] {\(\rho\)};

    \draw[Sprop, looparrow]
      (O) ++(270:\r) arc (270:450:\r)
      node[pos=0.5, right, xshift=6pt] {\(\rho\)};

  \end{feynman}
\end{tikzpicture}
\end{subfigure}
\hfill
\begin{subfigure}{0.48\textwidth}
\centering
\begin{tikzpicture}[scale=0.75]
  \begin{feynman}

    \def\r{1.2}
    \coordinate (O) at (0,0);

    \coordinate (vH) at ($(O)+(90:\r)$);      
    \coordinate (v1) at ($(O)+(210:\r)$);     
    \coordinate (v2) at ($(O)+(330:\r)$);    
    \node at ($(vH)-(0,0.25)$) {\(\lambda_{\Phi H}\)};
    \vertex [above left=1.8cm of vH]  (Hin)  {\(H\)};
    \vertex [above right=1.8cm of vH] (Hcin) {\(H\)};
    \vertex [below left=1.8cm of v1]  (a1)   {\(a_i\)};
    \vertex [below right=1.8cm of v2] (a2)   {\(a_i\)};

    \diagram*{
      (a1)   -- [scalar] (v1),
      (a2)   -- [scalar] (v2),
    };
    \draw[dashed,-Latex](Hin) -- ($(Hin)!0.55!(vH)$);
    \draw[dashed] ($(Hin)!0.55!(vH)$) -- (vH);
    
    \draw[dashed,-Latex] (vH) -- ($(vH)!0.55!(Hcin)$);
    \draw[dashed] ($(vH)!0.55!(Hcin)$) -- (Hcin);
    
    \tikzset{
      looparrow/.style={
        postaction={decorate},
        decoration={markings, mark=at position 0.55 with {\arrow{Latex}}}
      },
      Sprop/.style={dashed}
    }

    \draw[Sprop, looparrow]
      (O) ++(90:\r) arc (90:210:\r)
      node[pos=0.25, left, xshift=-5pt] {\(\rho\)};

    \draw[Sprop, looparrow]
      (O) ++(210:\r) arc (210:330:\r)
      node[pos=0.55, below] {\(a_i\)};

    \draw[Sprop, looparrow]
      (O) ++(330:\r) arc (330:450:\r)
      node[pos=0.75, above left, xshift=20pt, yshift=-6pt] {\(\rho\)};

  \end{feynman}
\end{tikzpicture}
\end{subfigure}

\caption{One-loop diagrams which contribute to the operator ${\cal O}_{Ha}$ when the heavy dark $\rho$ scalar is integrated out of the theory.  The lower vertices arise from the $\Phi$ kinetic energy term, while the upper vertices are from the $\lambda_{\Phi H}$ portal interaction.}
\label{fig:Higgs_loops}
\end{figure}

One-loop corrections also contribute at low energies to the $d=6$ universal DALP effective operator $\mathcal{O}_{Ha}$ through \cref {fig:Higgs_loops}. 
These one-loop contributions are subleading with respect to the tree-level one in \cref{tree-level-universal}, and we do not explore them further.

The contribution to $c_{Ha}$ obtained above will be present  in any construction involving the real triplet $\Phi$ undergoing spontaneous symmetry breaking, which includes all of the more general examples discussed below.

\subsection{The antisymmetric DALP operator \boldmath{$\mathcal{O}_{Ba}$} }
The antisymmetric DALP operator $\mathcal{O}_{Ba}$ in \cref{eq:DALP-LO-Lag} can be generated if a heavy dark field which carries hypercharge is present in the dark sector in addition to the hypercharge-neutral field $\Phi$.  We now explore the examples with additional heavy fermions or scalars carrying hypercharge.

\subsubsection{Heavy fermion triplet of $SO(3)$}\label{subsec:heavytriplet}
Consider the case with a heavy fermion field $\Psi_i$, which is a triplet of $SO(3)$ with hypercharge $Y_F$ and mass $\MF$.  The dark Lagrangian is then given by the scalar Lagrangian \cref{eq:Universal-loop} and in addition 
\begin{align}
\mathcal{L} &\supset \Bar{\Psi}(i\slashed{D}-\MF)\Psi
+ i y \epsilon_{ijk}\Bar{\Psi}_i\Psi_j \Phi_k \,,
\label{eq:LUV_2}
\end{align}
where the hypercharge gauge covariant derivative is $D_\mu \equiv \partial_\mu + i g_1 Y_F B_\mu$.
\begin{figure}
\begin{center}
\begin{tikzpicture}[scale=0.5]
\tikzset{photon/.style={decorate, decoration={snake, segment length=1.75mm, amplitude=0.75mm}}}
\draw (0,0) circle (2);
\draw[-latex] ([shift=(140:2)]0,0) arc (140:130:2);
\draw[-latex] ([shift=(50:2)]0,0) arc (50:40:2);
\draw[-latex] ([shift=(-40:2)]0,0) arc (-40:-50:2);
\draw[-latex] ([shift=(-130:2)]0,0) arc (-130:-140:2);
\draw[photon, line width = 0.5] (-4,0) -- (-2,0) node[midway, above] {$B_\mu$};
\draw[dashed] (90:4) -- (90:2) node[midway, right] {$\Phi$};
\draw[dashed] (-90:4) -- (-90:2) node[midway, right] {$\Phi$};
\draw[dashed] (0:4) -- (0:2) node[midway, above] {$\Phi$};
\node at (0,0) {$S,\Psi$};
\end{tikzpicture}
\end{center}
\caption{\label{fig:UV} One-loop graph which generates the operator ${\cal O}_{Ba}$. The external scalar lines are $\Phi$ fields, and the internal line is either the heavy scalar $S$ or the heavy fermion $\Psi$.  The external wiggly line is the $B_\mu$ gauge field.}
\end{figure}
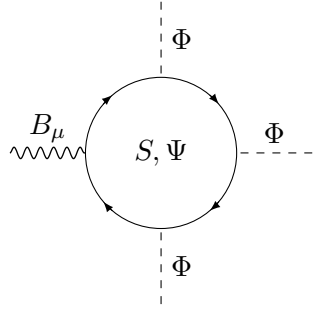
The tree-level contribution to $c_{Ha}$ in \cref{tree-level-universal} holds in this model, and furthermore the one-loop diagram in \cref{fig:UV} generates at low energies the antisymmetric DALP operator $\mathcal{O}_{Ba}$ in \cref{eq:DALP-LO-Lag}, with  
\begin{equation}
     \frac{c_{Ba}}{f_a^2}= \frac{ 22\,g_1\,Y_F \,y^3\, f_a}{45(4\pi)^2\, \MF^3}\,, 
\label{sol:LUV_2}
\end{equation}
where it has been assumed that the fermion masses  are large compared to all other scales in the UV theory, and we have treated the Yukawa contribution to the mass $y f_a$ as small compared to $M_F$. A more general analysis without this assumption is given in App.~\ref{sec:exact}.
  
\subsubsection{Heavy fermion doublet of $SO(3)$}
\label{subsec:heavydoublet}
Consider the case of a heavy dark fermion $\Psi_i$, which is a doublet of $SO(3)$ with hypercharge $Y_F$ and mass $M_F$. The dark Lagrangian is then given by the scalar Lagrangian \cref{eq:Universal-loop} plus the terms 
\begin{align}
\mathcal{L} &\supset \Bar{\Psi}(i\slashed{D}-\MF)\Psi
+ y\,\bar{\Psi}\,\tau^i \,\Psi  \, \Phi_i
,
\label{eq:LUV_3}
\end{align}
This example has a contribution to the universal DALP operator coefficient $c_{Ha}$ in \cref{tree-level-universal}, and in addition \cref{eq:LUV_3} generates the antisymmetric DALP  operator $\mathcal{O}_{Ba}$ in \cref{eq:DALP-LO-Lag} at low energies through the one-loop graph \cref{fig:UV}, resulting in  
\begin{equation}
     \frac{c_{Ba}}{f_a^2}= -  \frac{ 44\, g_1\, Y_F\, y^3\, f_a}{45(4\pi)^2 \MF^3}\,, 
\label{sol:LUV_3}
\end{equation}
where again the approximation in which $M_F$ is larger than any other scale of the UV theory has been taken, and we have expanded in the Yukawa coupling. Alike to the result for heavy fermion triplets in \cref{horrendous-1}, it is possible to use exact fermion mass eigenvalues to obtain a more complicated expression without expanding in $y$.

\subsubsection{Heavy scalar triplet}\label{subsec:heavyscalar}

The antisymmetric DALP operator is also generated in purely scalar UV theories.  Let us consider the example of a heavy complex scalar field $S_i$,  which is a triplet of $SO(3)$,  $S=(S^1,S^2,S^3)$, with hypercharge $Y_S$ and mass $\MS$. In the presence of both the $S$ field and the neutral real scalar triplet $\Phi$,   the UV dynamics is described by the Lagrangian 
\begin{align}
\label{eq-LUV}
\mathcal{L} &\supset (D_\mu S)^\dagger (D_\mu S) -\MS^2 (S^\dagger S) - \lambda_S (S^\dagger S)^2 
+\frac12 (D_\mu \Phi)^T (D_\mu \Phi) - \frac14 \lambda_\Phi (\Phi^T \Phi - f_a^2)^2 \nn \\
& + i \lam \epsilon_{ijk} S_i^* S_j  \Phi_k - \lambda_{SH} (S^\dagger S) (H^\dagger H) - \lambda_{\Phi H} (\Phi^T \Phi)(H^\dagger H)
-\lambda_{S\Phi} (S^\dagger S) (\Phi^T \Phi)
 \, ,
\end{align}
where $V(\Phi)$, given in \cref{Phi-pot}, induces $SO(3)$ spontaneous symmetry breaking,
and the second line of \cref{eq-LUV} contains the interactions between the heavy scalar triplets $S$ and $\Phi$ parametrized by $\mu$ and $\lambda_{S \Phi}$. The cubic interaction coupling $\lam$ has dimensions of mass, and preserves $C$ and $CP$.  Again, the $U(1)_Y$ gauge covariant derivative is $D_\mu =\partial_\mu + i g_1 Y_S B_\mu$. 

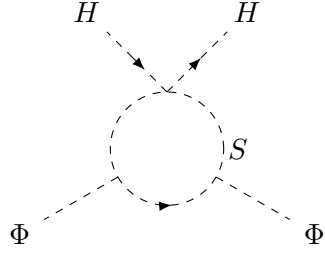
\begin{figure}
\begin{center}
\begin{tikzpicture}[scale=0.75]
\draw[dashed] (0,0) circle (1);
\draw[-Latex,dashed] (-95:1) arc (-95:-85:1);
\draw[dashed] (-30:1) -- (-30:2.5);
\draw[dashed] (210:1) -- (210:2.5);
\draw[dashed] (0,1) -- +(45:0.55);
\draw[dashed,-Latex](0,1)+(45:0.55) -- +(45:0.85);
\draw[dashed](0,1)+(45:0.85) -- +(45:1.5);
\draw[dashed](0,1) -- +(135:0.55);
\draw[dashed,-Latex](0,1)+(135:0.85) -- +(135:0.55);
\draw[dashed](0,1)+(135:0.85) -- +(135:1.5);
\draw (-30:3) node {$\Phi$};
\draw (210:3) node {$\Phi$};
\draw (0,1)+(45:2) node {$H$};
\draw (0,1)+(135:2)  node {$H$};
\draw (1.25,0) node {$S$};
\end{tikzpicture}
\end{center}
\caption{ \label{fig:higgsloop} One loop contribution to the axion-Higgs operator.}
\end{figure}

At energies much lower than $M_S$, the Lagrangian in \cref{eq-LUV} sources contributions mediated by the $S$ field to both the universal DALP operator ${\mathcal O}_{Ha}$ and the antisymmetric DALP operator ${\cal O}_{Ba}$.  For $\mathcal{O}_{Ha}$, both the  tree-level $\rho$ exchange contribution  discussed above (see \cref{tree-level-universal}) and the one-loop contribution depicted in \cref{fig:higgsloop} are present, 
\begin{equation}
    \frac{c_{Ha}}{f_a^2}= -\frac{\lambda_{\Phi H}}{\lambda_\Phi f_a^2}\,-\frac{\lambda_{SH}\, \mu^2}{ 6 \left(4\pi\right)^2 \MS^4}\,.
\end{equation}
For the antisymmetric operator $\mathcal{O}_{Ba}$, the first contribution appears at one-loop order from the Feynman diagram in \cref{fig:UV}. Evaluating the graph at low energies,  where the external momenta are small compared to the heavy scalar mass $\MS$, leads to the effective interaction
\begin{align}
\mathcal{L} &= \frac{1}{90(4 \pi)^2}  \frac{g_1 Y_S \lam^3}{\MS^6}   \left[  B_{\mu \nu}\ \epsilon_{ijk} \ \Phi_i\, \partial_\mu \Phi_j\, \partial_\nu \Phi_k \right] + \mathcal{O} \left( \frac{1}{\MS^7} \right) \, .
\label{3.2}
\end{align}
In the broken phase,
\begin{align}
\epsilon_{ijk} \ \Phi_i\, \partial_\mu \Phi_j\, \partial_\nu \Phi_k = f_a^3 \epsilon_{ij} u_\mu^i u_\nu^j \,,
\label{3.9}
\end{align}
so that \cref{3.2} generates the $\mathcal{O}_{Ba}$ operator, with
\begin{align}
    \frac{c_{Ba}}{f_a^2}= \frac{ g_1\, Y_S\,  f_a\, \mu^3 }{90(4\pi)^2 \MS^6} \,.
    \label{6.18}
\end{align}
In \cref{fig:UV}, we have implicitly assumed that $\mu \ll \MS$, thus allowing us to treat the $\mu$ interaction term  in \cref{eq-LUV} as a perturbation.  
The more general result without this assumption is given in App.~\ref{sec:exact}.

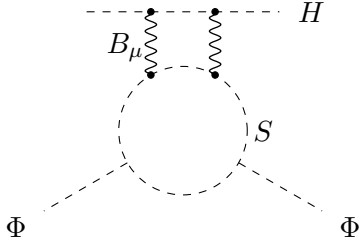
\begin{figure}
\begin{center}\begin{tikzpicture}[scale=0.85]

\tikzset{photon/.style={decorate, decoration={snake, segment length=1.75mm, amplitude=0.75mm}}}

\draw[dashed]  (0,0) circle (1);
\draw[photon, line width = 0.5] (60:1) -- +(0,1);
\draw[photon, line width = 0.5] (120:1) -- +(0,1);
\draw[dashed] (-1.5,1.85) -- (1.5,1.85);
\filldraw (0.5,1.85) circle (0.05);
\filldraw (-0.5,1.85) circle (0.05);
\filldraw (60:1) circle (0.05);
\filldraw (120:1) circle (0.05);
\draw (2,1.85) node {$H$};
\draw (-0.9,1.3) node {$B_\mu$};
\draw[dashed] (-30:1) -- (-30:2.5);
\draw[dashed] (210:1) -- (210:2.5);
\draw (-30:3) node {$\Phi$};
\draw (210:3) node {$\Phi$};
\draw (1.25,0) node {$S$};

\end{tikzpicture}
\end{center}
\caption{\label{fig:2loop} Two-loop contribution to the ${\mathcal O}_{Ha}$ operator. }
\end{figure}

The UV Lagrangian \cref{eq-LUV} generates $\mathcal{O}_{Ha}$ at tree-level and $\mathcal{O}_{Ba}$ at one-loop. However, one can imagine a scenario where the $G/H$ fields of $\Phi$ are decoupled, except for the complex scalar $S$ that couples to hypercharge.  The coupling between the $G/H$ and SM sectors is due to the interaction proportional to $\mu$, which has  dimension of mass.  This interaction between the two sectors is soft, and in consequence the theory does not induce  $\lambda_{SH}$ and $\lambda_{\Phi H}$ couplings, and it is consistent to set them to zero since these couplings are not required for renormalizability of the theory.  In this case, $\mathcal{O}_{Ha}$ is not generated at tree-level. The first contribution to $O_{Ha}$ arises then instead from the two-loop graph in \cref{fig:2loop},
which generates a contribution to $\lambda_{\Phi H}$ of order $g_1^4/(4\pi)^4 \mu^2/\MS^2$. In the heavy fermion examples, the analogous graph with $S$ replaced by $\Psi$ is UV divergent, with a divergence of order $\lambda_{\Phi H} \sim g_1^4 y^2/(4\pi)^4 \ 1/\epsilon$.  In summary, in spite of their loop-vs-tree nature, scenarios in which the coefficient of the  hypercharge operator $\mathcal{O}_{Ba}$ is comparable or larger than that of the DALP-Higgs interaction $\mathcal{O}_{Ha}$ are plausible and technically natural; both  couplings should be searched for on equal footing. 

\subsection{Generalizations}\label{sec:uvcompletions2}

The UV models discussed so far can be generalized to arbitrary $G \to H$ symmetry breaking patterns. The $\mathcal{O}_{Ha}$ operator can easily be generated at tree-level by coupling dark sector scalars to the SM using the Higgs portal $(H^\dagger H)(\Phi^\dagger \Phi)$, where $\Phi$ is a dark sector scalar,  as in \cref{eq:Universal-loop}. Generating $\mathcal{O}_{Ba}$ is more difficult, so we focus on this operator here.

There is a class of UV models which generate $\mathcal{O}_{Ba}$, where the symmetry breaking pattern $G\to H$ is generated by the VEV of a real scalar $\Phi$ in the adjoint of $G$. This case corresponds to a large class of models, covering breaking patterns where $H$ is a regular subgroup of $G$ of maximal rank: it includes the generalizations of examples~(a) and~(b) in \cref{sec:examples} and the other families listed in \cref{subsubsec:OtherSBBpatterns}. The low-energy interaction $\mathcal{O}_{Ba}$ is given by integrating out a heavy scalar or heavy fermion, and we give one example of each.

\paragraph{(i) Integrating out a heavy scalar:} The UV theory has a real scalar $\Phi_A$ in the adjoint representation of $G$, and a complex scalar $S_A$ in the adjoint representation of $G$ with hypercharge $Y_S$. $S$ is the mediator for $B_{\mu \nu}$ between the dark sector and the SM. The relevant interactions are
\begin{align}
\mathcal{L} &= (D_\mu S)^\dagger (D_\mu S) -\MS^2 (S^\dagger S)  + i \lam f_{ABC} S^*_A S_B \Phi_C \,.
\label{3.10}
\end{align}
The cubic interaction $\lam$ has dimensions of mass, and preserves $C$ and $CP$.

The graph in \cref{fig:UV} gives the effective interaction at low energies
\begin{align}
\mathcal{L} &= \frac{1}{180  (4\pi)^2}  \frac{g_1 Y_S \lam^3 C_{Ad}}{\MS^6}   \left[  B_{\mu \nu}\ f_{ABC} \ \Phi_A\, \partial_\mu \Phi_B\, \partial_\nu \Phi_C \right] + \mathcal{O} \left( \frac{1}{\MS^7} \right) \, ,
\label{3.11}
\end{align}
where $C_\text{Ad}$ is the Casimir of the adjoint, $f_{ABC} f_{ABD}= C_\text{Ad} \, \delta_{CD}$. \Cref{3.11} reduces to the result for the scalar $SO(3)$ example \cref{6.18} using $C_\text{Ad}=2$.

\paragraph{(ii) Integrating out a heavy fermion:} The UV theory has a real scalar $\Phi_A$ in the adjoint representation of $G$ and a fermion $\Psi$ in a representation $\mathbf{{R_F}}$ of $G$ with hypercharge $Y_F$, and interaction Lagrangian is
\begin{align}
\mathcal{L} &= \overline \Psi\, i \slashed{D}\, \Psi - \MF \overline \Psi \Psi  + y \left( \overline \Psi\, t^A P_{R}\, \Psi \right) \, \Phi_A +  y^* \left( \overline \Psi\,  t^A P_L\,  \Psi \right) \, \Phi_A \,,
\label{3.3}
\end{align}
which violates $P$ and $CP$ if $y \not = y^*$.
Integrating out the fermion, the graph in \cref{fig:UV} gives the effective interaction
\begin{align}
\mathcal{L} &= \frac{1}{90(4 \pi)^2}  \frac{g_1 Y_F T_{R_F}}{\MF^3}   \left[  \left(C B_{\mu \nu} + i \widetilde C \widetilde B_{\mu \nu} \right) f_{ABC} \ \Phi_A\, \partial_\mu \Phi_B\, \partial_\nu \Phi_C \right] +  \mathcal{O} \left( \frac{1}{\MF^4} \right) \, ,
\label{3.4}
\end{align}
where $T_{R_F}$ is the Dynkin index of $\mathbf{{R_F}}$, defined as $T_{R_F} \delta_{A B}\equiv \operatorname{Tr}\left(t_{R_F}^A t_{R_F}^B\right)$, with 
\ytableausetup{boxsize=0.5em}
$T(\ydiagram{1})=1/2$, and
\begin{align}
C&= \left[ y+y^*\right]\left[ y^2 + (y^*)^2 - 13 y y^* \right] \,, &
\widetilde C &= 15 y y^*  \left[ y-y^*\right]\,.
\label{3.5}
\end{align}
The $\widetilde B_{\mu \nu}$ operator violates $P$ and $CP$. The result \cref{3.4} gives \cref{sol:LUV_2} and \cref{sol:LUV_3} for the $SO(3) \to SO(2)$ case.

At energies below the mass scale $\MS$, $\MF$ of the heavy particles, the adjoint field $\Phi_A$ develops a vacuum expectation value $\vev{\Phi_A}$ which breaks the symmetry $G$ to a subgroup $H$, and gives rise to the ALP Goldstone bosons. The VEV $\vev{\Phi_A}$ can be diagonalized, so that it is in the Cartan subalgebra of $G$. The Goldstone boson fluctuations are parametrized by
 \begin{align}
\Phi &= \xi \vev{\Phi}\xi^{-1} \, , & \Phi &= \Phi_A t_A \, ,
\label{3.6}
\end{align}
and
\begin{align}
f_{ABC} \ \Phi_A\, \partial^\mu \Phi_B\, \partial^\nu \Phi_C & \to  2 i u_i^\mu u_j^\nu  \tr \left\{  \vev{\Phi}  \left[ [X_i ,\vev{\Phi}] , [X_j, \vev{\Phi}] \right]
\right\}  \, .
\label{3.7}
\end{align}
At LO in the Goldstone fields, $u^i_\mu \to \partial_\mu a^i/f_a$, and so \cref{3.7} gives
\begin{align}
   B_{\mu\nu }u_i^\mu u_j^\nu  A_{ij}
  \;\to\;
  \frac{1}{f_a^2}B_{\mu\nu }\,\partial^\mu a_i\,\partial^\nu a_j A_{ij} \;\propto\; \mathcal{O}_{Ba}\,,
\end{align}
where $A_{ij} \propto\, i\,\tr \left\{ \langle\Phi\rangle\bigl[[X^i,\langle\Phi\rangle],[X^j,\langle\Phi\rangle]\bigr] \right\}$ constructively defines the antisymmetric $H$-invariant tensor required to build the operator $\mathcal{O}_{Ba}$.

 All  examples discussed in this section are UV completions in the perturbative regime.  It would be interesting to explore whether, instead, the breaking patterns involved can be realized dynamically in a confining theory with chiral symmetry breaking. In that case, the composite pGB would still couple to the SM through the operators discussed in this paper starting with $\mathcal{O}_{Ha}$ and $\mathcal{O}_{Ba}$  in Eq.~(\ref{eq:DALP-LO-Lag}), but the analysis of the radial mode $\rho$ would be qualitatively modified (a plethora of resonances would then contribute). In particular, the bounds from Higgs-exotic scalar mixing within our weakly coupled UV completion in \cref{subsec:UniversalDALPOp} would no longer apply directly, potentially decorrelating the effective DALP operators from those bounds.

\section{Conclusions}\label{sec:conclusions}

The standard ALP framework implicitly assumes that ALPs are neutral, not only under the Standard Model gauge group, but also under any potential symmetry of the dark sector. In this work, we have explored the consequences of relaxing this assumption and allowing ALPs to carry conserved dark charges. We find that this simple change qualitatively modifies the structure of the low-energy EFT and, consequently, its associated phenomenology. 

In this paper, we studied a dark ALP scenario where the ALPs are pseudo-Goldstone bosons arising from a dark sector symmetry breaking $G \to H$, and they  transform non-trivially under the unbroken symmetry $H$. We named these darkly-charged ALPs DALPs.  The new feature in this scenario is that single DALP interactions with the SM, such as the usual $d=5$ ALP couplings, are forbidden. The leading DALP interactions are then $d=6$, and given by two possible EFT operators: $\mathcal{O}_{Ha}$, which couples the DALPS to the Higgs field, and $\mathcal{O}_{Ba}$ which couples the DALPs to the  $U(1)_Y$ field-strength $B_{\mu \nu}$. The $\mathcal{O}_{Ha}$ operator is universal, while the $\mathcal{O}_{Ba}$ is sensitive to the symmetry breaking pattern and  it can thus discriminate among classes of dark symmetries. 

We have analyzed the DALP couplings to the SM using the CCWZ formalism to determine the full non-linear form of the interactions, and classified the allowed chiral operators up to $d=9$.  Some group-theoretic conditions are derived for the existence of DALP couplings, which simplifies the search for DALP symmetry breaking patterns.

We have explored the phenomenology of the leading order $d=6$ DALP interactions in the case where the DALPs are degenerate, for simplicity. Degeneracy naturally arises if $G$ is explicitly broken while preserving $H$. The invisible widths of the Higgs and $Z$ constrain the coefficients of $\mathcal{O}_{Ha}$ and $\mathcal{O}_{Ba}$, respectively. We have examined the constraints from monojet signals, as well as from supernova cooling bounds and direct searches.

DALPs are also good dark matter candidates, and the correct relic abundance can be obtained through several mechanisms. The freeze-out mechanism is excluded, except in tiny regions of parameter space, due to the Higgs and $Z$ resonances. Instead, obtaining the correct relic abundance via the freeze-in scenario is possible, and we have mapped out the allowed region in parameter space. Finally, DALPs can also account for the dark matter through misalignment, which allows for large DALP masses reaching up to hundreds of MeV.

While the heart of the paper relies on an effective field theory analysis and thus is independent of specific UV complete models, we also have presented explicit examples of perturbative UV theories which can generate the $d=6$ DALP couplings. Specifically, we considered UV theories with heavy scalar or with heavy fermions based on the symmetry breaking pattern $SO(3) \to SO(2)$, as well as generalizations to arbitrary $G \to H$ breaking.

In summary, given the complexity and non-trivial quantum numbers of the visible world, it is natural to consider the possibility that the dark sector also carries complex quantum numbers, and that conserved  dark charges exist in Nature. From this perspective, DALPs should not be regarded as a special corner of ALP parameter space, but as a qualitatively distinct possibility. While we focused on the simplest EFT realization --identifying the expected leading operators and signals-- the framework introduced here opens many avenues for future exploration. In particular, it would be interesting to study the impact of the higher-order interactions predicted by the EFT, and  the richer phenomenology expected in scenarios with non-degenerate DALP spectra, as well as possible dark gauge interactions if subgroups of $G$ and $H$ are gauged. The DALP scenario opens vast new territory beyond that of the usual chargeless ALPs, and requires new strategy to search for the leading interactions of DALPs with SM particles.

\section{Acknowledgments}\label{sec:ack}

We thank Itay Bloch, Valerie Domcke, Victor Enguita-Vileta, Yohei Ema, Benjam\'\i n Grinstein, David B. Kaplan, Maria Ramos, Tracy Slatyer, Anders Eller Thomsen and Felix Yu for useful discussions. This work was partially supported by the U.S. Department of Energy (DOE) award number DE-SC0009919. It also has received support from the European Union’s Horizon 2020 research and innovation programme under the Marie Skłodowska-Curie grant agreement No 101086085 - ASYMMETRY, and from the Spanish Research Agency (Agencia Estatal de Investigaci\'on) through Grant IFT Centro de Excelencia Severo Ochoa No CEX2020-001007-S and grant PID2022-137127NB-I00, funded through MCIN/AEI/10.13039/501100011033, and by “European Union NextGenerationEU/PRTR'' and by ERDF/EU. P.Q. acknowledges support by the European Union's Horizon 2020 research and innovation programme under the Marie Sk\l odowska-Curie Postdoctoral Fellowship grant agreement No 101207780 - AxionCount, and thanks the UCSD Physics Department for its hospitality. N.M.A. acknowledges support from the Spanish Ministry of Science, Innovation and Universities (MICIU) through the predoctoral fellowship FPU/00948.

\begin{appendix}

\section{Proofs of the refined group-theoretic conditions}
\label{app:proofs}

In this appendix, we provide proofs of the conditions discussed in \cref{Subsec:refining2}.

\subsection{Proof of equivalent condition for {\bf [C1]} in \cref{Subsec:refining2} }
\label{app:proofC1}

\noindent\textbf{Proposition}: Condition {\bf [C1]} holds if and only if every $U(1)$ subgroup of $G$ that commutes with all of $H$ is contained in $H$. 

\textit{($\Rightarrow$)} Assume there exists a $U(1)\subset G$ commuting with all the elements in $H$ but not contained in $H$. Its generator is therefore a broken generator. Since $[Y,T_\alpha]=0$ for all $T_\alpha\in\adj(H)$, it is $H$-invariant, so $\RGB\supset\I_H$.\hfill$\square$

\textit{($\Leftarrow$)} Assume $\RGB\supset\I_H$. Then some broken generator $Y\notin H$ satisfies $[Y,T_\alpha]=0$ for all $T_\alpha\in\adj(H)$. Then $e^{i\theta Y}$ is a $U(1)\subset G$ that commutes with all of $H$ but is not contained in $H$.

\subsection{Sufficient condition for {\bf[C2]}}
\label{app:centerproofC2}

\noindent\textbf{Proposition}: If the center of $H$ is not contained in the center of $G$, $Z(H)\not\subset Z(G)$, and the quotient
\begin{equation}\label{eq:quotientK}
  K \;\equiv\; Z(H)\,\big/\,\big(Z(H)\cap Z(G)\big)
\end{equation}
contains an element of order $n\geq 3$, then $\RGB$ contains a pair
of complex conjugate irreps $\mathbf{c}\oplus\bar{\mathbf{c}}$, and therefore
condition~{\bf [C2]} is satisfied: $(\RGB\otimes \RGB)_A\supset\I_H$.

\medskip
\textit{Proof.} Let us pick $z\in K$ of order $k\geq 3$.  Since $z\in Z(H)$, it acts trivially on $\adj(H)$, but since $z\notin Z(G)$ it acts non-trivially on $\adj(G)$. From $\adj(G)|_H=\adj(H)\oplus\RGB$ this non-trivial action must lie in $\RGB$. Since $k\geq 3$, at least one eigenvalue on $\RGB$ is a $k$-th root of unity, thus complex. Since $\adj(G)$ is real its conjugate irrep also appears in $\RGB$, implying that $\RGB\supset\mathbf{c}\oplus\bar{\mathbf{c}}$ where $\mathbf{c}$  is a complex irrep and thus $(\RGB\otimes\RGB)_A\supset\I_H$.\hfill$\square$

\medskip\noindent\textbf{Corollary.} As a special case, we get the sufficient condition discussed in \cref{Subsec:refining2} --- If $H\simeq H^\prime \times U(1)$ with the $U(1)$ factor not a direct factor of $G$, then condition {\bf [C2]} is satisfied.

\medskip\noindent\textit{Proof.} Since $U(1)\subseteq Z(H)$ but $U(1)\not\subseteq Z(G)$, $K$ contains a copy of $U(1)$ and in particular complex elements so the condition {\bf [C2]} is satisfied.\hfill$\square$

\section{Generalization of Matching Calculations}\label{sec:exact}

In this appendix, we generalize the results of \cref{subsec:heavytriplet} and  \cref{subsec:heavyscalar} without expanding in the Yukawa coupling $y$ and the dimensionful coupling $\mu$, respectively. All results obtained in this appendix were obtained using the \texttt{Matchete} package for Wolfram Mathematica~\cite{Fuentes-Martin:2022jrf}.

The fermions masses for the example in \cref{subsec:heavytriplet} are $M_F$ and $M_{\pm}=\MF \pm y f_a$. Assuming these fermions masses are large compared to other scales in the problem, then integrating the heavy fermions out of the theory gives the generalization of \cref{sol:LUV_2},
\begin{align}
    \frac{c_{Ba}}{f_a^2}=&\frac{2 g_1 {Y}_F}{3 (4 \pi)^2(f_a^2 y^2-4  \MF^2)^2 y^2 f_a^4}\bigg[ ( 12 \MF^6 - 13 \MF^4 f_a^2 y^2 -f_a^6 y^6 +2 \MF^2 f_a^4 y^4) L_-    \nonumber\\ 
    &+(12 f_a y \MF^5 + 3 f_a^5 y^5 \MF)(4 + L_+) - 6 f_a^3 y^3 \MF^3 (10+L_+)\bigg],
    \label{horrendous-1}
\end{align}
where we have introduced the auxiliary function
\begin{align}
    L_{\pm}\equiv \ln \frac{\MF^2}{(\MF+ f_a y)^2} \pm  \ln \frac{\MF^2}{(\MF - f_a y)^2} \,.
\end{align}
 It can be easily checked that the apparent divergence of the overall coefficient for $M_F= yf_a/2$ is cancelled by the rest of the expression, leading to a finite contribution
 \begin{align}
     \frac{c_{Ba}}{f_a^2}\to \frac{g_1 Y_F}{48 (4 \pi)^2 f_a^2}(12+\ln (3)).
 \end{align}
We obtained Eq.~\eqref{horrendous-1} by integrating out all heavy particles at the same RG scale $M_F$, which requires that $M_+\sim M_-\sim M_F$. There is a logarithmic divergence in the limit $M_- \to 0$. To understand this regime, we may consider the limiting case $M_-=0$, i.e. $M_F=y f_a$.\footnote{Note that this does not correspond to a strictly massless lightest particle, as its mass is subject to a loop-level correction $\delta M_- \sim y^2 M_f/(4 \pi)^2.$} Integrating out only the heavy particles (keeping the lightest fermion in the spectrum), we again obtain a finite Wilson coefficient,
\begin{align}
    \frac{c_{Ba}}{f_a^2}=& \frac{2 g_1 Y_F}{3 (4 \pi)^2 f_a^2} \ln (4)\,,
\end{align}
where the low energy theory now has an additional massless dark fermion.

Similarly, we can compute the generalization of the matching \cref{sol:LUV_3} for the example in \cref{subsec:heavydoublet}. Here, the masses of the heavy fermions are given by $M_{\pm}=M_F\pm f_a y$. In terms of these masses, we find for the Wilson coefficient
\begin{align}
    \frac{c_{Ba}}{f_a^2}=\frac{g_1 Y_F}{96 (4 \pi)^2 f_a^4 y^2 M_F^2}\bigg[& 6 M_+ M_-^3 \left(2- L_{\pm} \right) +2 M_+^2 M_-^2 L_\pm -6 M_- M_+^3 (2+L_\pm) \nonumber \\ 
    &- M_-^4 (3-2L_\pm) + M_+^4 (3+2 L_\pm) \bigg],
\end{align}
with the auxiliary function $L_{\pm}=\ln (M_-^2/M_+^2)$. 

Similar to the previous model, this expression develops a logarithmic divergence in the limit $M_{\pm}\to 0$ as a consequence of integrating out the two particles at the same RG scale despite their parametrically different masses. Setting $M_-=0$ and only integrating out the heavy particle $\Psi_2$, we find for the consistent limiting value
\begin{align}
    \frac{c_{Ba}}{f_a^2}=\frac{g_1 Y_F}{6 (4 \pi)^2 f_a^2}( 3-2 \ln 4)\, .
\end{align}
Finally, we repeat our calculation for \cref{6.18} for the example in \cref{subsec:heavyscalar}. The eigenvalues of the mass matrix of $S$ are $M_S^2$ and $M_\pm^2=\MS^2\pm \mu f_a$.  From the expression for the lightest component
$M_-$, it follows that  $\mu f_a<M_S^2$ to avoid spontaneous symmetry breaking of the hypercharge symmetry via the VEV of $S$.  This condition does, however, allow for regimes in which $\mu\sim \MS$. For those cases, the matching can be performed by rotating the $S$ fields into mass eigenstates. Doing this, we find the expression 
\begin{align}
    \frac{c_{Ba}}{f_a^2}=-\frac{g_1\,Y_S}{6(4\pi)^2 f_a^4\, \mu^2 }\, \bigg\{ & \left( 6 \MS^4 + f_a^2\,  \mu^2\right)\left[ \log\left(1 - \frac{ f_a \, \mu}{\MS^2}\right) - \log\left(1 + \frac{ f_a\,  \mu}{\MS^2}\right) \right]  \nonumber \\
     &- 6 f_a \, \mu \MS^2 \left[ \log\left( 1 - \frac{ f_a \, \mu}{\MS^2}\right) + \log\left( 1 + \frac{ f_a \, \mu}{\MS^2}\right) -2  \right] \biggr\}  ,
     \label{horrendous-2}
\end{align}
which agrees with \cref{6.18} to LO in an expansion in $\mu/\MS$.

This form was again obtained by simultaneously integrating out all heavy particles at the matching scale $M_S$, manifesting in a logarithmic divergence if $M_{\pm}\to 0$. By instead integrating out only the heavy particles while keeping the lightest particle in the spectrum, we find in the limit $M_S^2=\mu f_a$,
\begin{align}
    \frac{c_{Ba}}{f_a^2}=-\frac{g_1 Y_S}{6 (4 \pi)^2 f_a^2}(\ln (8192)-12)\simeq \frac{g_1 Y_S}{2 (4 \pi)^2 f_a^2}\, .
\end{align}

\end{appendix}

\bibliographystyle{JHEP}
\bibliography{BiblioDGB}

\end{document}